\documentclass[fleqn,usenatbib]{mnras}

\usepackage{newtxtext,newtxmath}

\usepackage[T1]{fontenc}
\usepackage{ae,aecompl}

\usepackage{graphicx}	
\usepackage{amsmath}	
\usepackage{amssymb}	
\usepackage{rotating}
\usepackage{pdflscape}
\newcommand{\kms}{km~s$^{-1}$}
\newcommand{\cmcub}{cm$^{-3}$}
\newcommand{\oiii}{[OIII]$\lambda5007$}
\newcommand{\hb}{H$\beta$}
\newcommand{\halpha}{H$\alpha$}

\title[The Abell 2667 BCG: properties from MUSE]{Inquiring into the nature of the Abell 2667 Brightest Cluster Galaxy: physical properties from MUSE}
\author[E. Iani et al.]{
E. Iani$^{1,2}$\thanks{E-mail: edoardo.iani@phd.unipd.it},
G. Rodighiero$^{1,3}$,
J. Fritz$^4$,
G. Cresci$^5$,
C. Mancini$^{1,3}$,
P. Tozzi$^5$,
\newauthor
~L. Rodr\'iguez-Mu\~noz$^1$,
P. Rosati$^6$,
G. B. Caminha$^7$,
A. Zanella$^{2}$,
S. Berta$^8$,
\newauthor
~P. Cassata$^{1}$,
A. Concas$^{9}$,
A. Enia$^1$,
D. Fadda$^{10}$,
A. Franceschini$^1$,
A. Liu$^{5}$,
\newauthor
~A. Mercurio$^{11}$,
L. Morselli$^{9}$,
P. G. P\'erez-Gonz\'alez$^{12,13}$,
P. Popesso$^{9}$,
\newauthor
~G. Sabatini$^{14,15,16}$,
J. Vernet$^{2}$,
R. J. van Weeren$^{17}$
\\
\\
Affiliations are listed at the end of the paper
}

\date{Accepted XXX. Received YYY; in original form ZZZ}

\pubyear{2015}

\begin{document}
\label{firstpage}
\pagerange{\pageref{firstpage}--\pageref{lastpage}}
\maketitle

\begin{abstract}
Based on HST and MUSE data, we probe the stellar and gas properties (i.e. kinematics, stellar mass, star formation rate) of the radio-loud brightest cluster galaxy (BCG) located at the centre of the X-ray luminous cool core cluster Abell 2667 ($z = 0.2343$).
The bi-dimensional modelling of the BCG surface brightness profile reveals the presence of a complex system of substructures extending all around the galaxy.
Clumps of different size and shape plunged into a more diffuse component constitute these substructures, whose intense `blue' optical colour hints to the presence of a young stellar population.
Our results depict the BCG as a massive (M$_\star \simeq 1.38 \times 10^{11}~\text{M}_\odot$) dispersion-supported spheroid ($\Delta \text{v}_\star \leq 150$ \kms, $\sigma_0 \sim 216$ \kms) hosting an active supermassive black hole (M$_{SMBH} \simeq 3.8\times 10^{9}~\text{M}_\odot$) whose optical features are typical of low ionisation nuclear emission line regions. 
Although the velocity pattern of the stars in the BCG is irregular, the stellar kinematics in the regions of the clumps show a positive velocity of $\sim 100$ \kms, similarly to the gas component.
An analysis of the mechanism giving rise to the observed lines in the clumps through empirical diagnostic diagrams points out that the emission is \textit{composite}, suggesting the contribution from both star formation and AGN.
We conclude our analysis describing how scenarios of both \textit{chaotic cold accretion} and \textit{merging} with a gas-rich disc galaxy can efficaciously explain the phenomena the BCG is undergoing.
\end{abstract}




\section{Introduction}
Brightest cluster galaxies (BCGs) are the largest, most massive and optically luminous galaxies in the Universe.
They sit almost at rest at the bottom of clusters potential wells and close to the peak of the thermal X-ray emission originated by the hydrostatic cooling of the hot ($10^7-10^8$~K) intracluster medium (ICM).
Based on their optical morphology and their `red' optical/near-infrared colours that suggest relatively old stellar populations and little ongoing star formation activity (e.g. \citealt{1998ApJ...502..141D}; \citealt{2012ApJ...759...43T}; \citealt{2014ApJ...789..134B}; \citealt{2015MNRAS.453.4444Z}), BCGs are typically classified as giant ellipticals or cD galaxies.
However, they distinguish themselves from cluster ellipticals because of different surface brightness profiles (i.e. shallower surface brightness and higher central velocity dispersion e.g. \citealt{1976ApJ...209..693O}; \citealt{1986ApJS...60..603S}, \citealt{1998ApJ...502..141D}), larger masses and luminosities (BCGs are often found to be a few times brighter than the second and third brightest galaxies in a cluster).
Additionally, BCGs have been shown to lie off the standard scaling relations of early-type galaxies (e.g. \citealt{2007AJ....133.1954B}; \citealt{2007ApJ...670..249L}; \citealt{2007MNRAS.379..867V}; \citealt{2009MNRAS.395.1491B}) with luminosities (or stellar masses) significantly above the prediction of the standard Faber-Jackson relation (e.g. \citealt{2014ApJ...797...82L}).
For these reasons, BCGs do not appear to represent the high-mass end of the luminosity function of either cluster ellipticals (e.g. \citealt{1977ApJ...212..311T}, \citealt{1984ARA&A..22..185D}) or bright galaxies in general (\citealt{2001MNRAS.322..625B}), defining in this way a galaxy class on their own (e.g. \citealt{1984ARA&A..22..185D}; \citealt{2007MNRAS.375....2D}; \citealt{2017MNRAS.469.1259B}).

As a result of their peculiar location within clusters, BCGs can show exceptional properties over the whole electromagnetic spectrum.
Hence, a profusion of studies probing their nature has been carried out from X-rays to the radio domain (e.g. \citealt{2000ApJ...532L.105B}; 
\citealt{2010ApJ...718...23S}; 
\citealt{2012MNRAS.427..550L}; 
\citealt{2014MNRAS.442.2048D}; 
\citealt{2015ApJ...814...96W}). 
The results of these works show that BCGs with bright mid- and far-infrared emission likely host star formation as well as active galactic nuclei (AGNs).
In this case, the infrared emission is the consequence of the absorption and re-emission by dust of UV light generated by either young stars or the hard radiation field of AGNs.
Both on-going star formation and AGN activity induce BCGs to be detected in the radio domain. 
In this regard, studies on the fractional radio luminosity function of elliptical galaxies (e.g. \citealt{1977A&A....57...41A}; \citealt{1996AJ....112....9L}; \citealt{2007MNRAS.375..931M}; \citealt{2010A&A...511A...1B}) have shown that the probability of a galaxy hosting a radio-loud source strictly depends on its visual magnitude.
Specifically, for a given galaxy radio power, the radio luminosity function increases depending on the galaxy mass and the crowdedness of its environment.
Therefore BCGs are by definition the galaxies with the highest probability to host an AGN (e.g. \citealt{2007MNRAS.379..894B}; \citealt{2007MNRAS.379..867V}; \citealt{2009A&A...501..835M}).
Indeed, a radio survey of BCGs residing in CLASH (i.e., the Cluster Lensing And Supernova survey with \textit{Hubble}, \citealt{2012ApJS..199...25P}) clusters show that BCGs frequently host a radio galaxy, with the radio emission at least 10 times more powerful than expected from measured star formation (e.g.~\citealt{2018ApJ...853..100Y}). 
If the AGN is sufficiently unobscured and radiatively efficient, it is detected in the X-rays (e.g. \citealt{2018ApJ...859...65Y}) with its optical counterpart featuring emission line ratios typical of low ionisation nuclear emission regions (LINER; e.g. \citealt{1989ApJ...338...48H}; \citealt{1999MNRAS.306..857C}). 
Ultimately, evidence of a multiphase ICM surrounding BCG cores has been unveiled by the X-ray (e.g. \citealt{1994ARA&A..32..277F}), in the UV via the OVI line \citep{2001ApJ...553L.125B}, the extended Ly$\alpha$ \citep{1992ApJ...391..608H} and far-UV continuum (e.g. \citealt{2004ApJ...612..131O}) emissions, in the optical, predominantly via the \halpha$+$[NII]~emission (e.g. \citealt{1989ApJ...338...48H}; \citealt{1999MNRAS.306..857C}; \citealt{2011ApJ...742L..35M}), in the near-IR and mid-IR via the roto-vibrational (\citealt{2000ApJ...545..670D}, \citealt{2002MNRAS.337...49E}) and the pure rotational H$_2$ lines (\citealt{2006ApJ...652L..21E}; \citealt{2007MNRAS.382.1246J}, \citealt{2011ApJ...732...40D}).

In the last decades, a plethora of studies have investigated both observationally and theoretically the mechanisms at the origin of BCG formation. 
The results obtained in these works invoke either \textit{cooling flows} (\citealt{1976ApJ...208..646S}; \citealt{1977ApJ...215..723C}; \citealt{1977MNRAS.180..479F}; \citealt{1994ARA&A..32..277F}) or \textit{galactic cannibalism} (\citealt{1975ApJ...202L.113O}; \citealt{1976MNRAS.177..717W}; \citealt{1984ApJ...276..413M}; \citealt{1985ApJ...289...18M}) to explain the stellar mass growth of these galaxies.
Cooling flows are streams of subsonic, pressure-driven cold ICM (e.g. \citealt{2010ApJ...719.1619O}) sinking towards the centre of the clusters potential wells. 
The physical conditions giving rise to cooling flows depend on the properties of the ICM and the hosting cluster.
In particular, if the central ($r\lesssim 10 - 100$~kpc, $\sim 10$\% of the virial radius, e.g. \citealt{1997MNRAS.292..419W}; \citealt{2010A&A...513A..37H}; \citealt{2017ApJ...843...28M}) density of the ICM in dynamically relaxed clusters is $0.1~\text{cm}^{-3}$ and its temperature (as obtained from X-ray observations) is $\sim 10^7$ K, the ICM ought to cool via thermal bremmstrahlung in a timescale significantly shorter than the Hubble time (e.g. \citealt{1998MNRAS.298..416P}; \citealt{2004MNRAS.347.1130V}).
These cosmologically rapidly cooling regions, are referred as `cool cores' and today are believed to characterise roughly a third of all galaxy clusters (e.g. \citealt{2018ApJ...858...45M}), the so-called cool core clusters.
According to their properties, cooling flows could drive to the BCG region star-forming gas at huge rates.
As a matter of fact, if the total mass of the ICM in cool cores is integrated and divided by the cooling time, the theoretical isobaric ICM cooling rate for a massive galaxy cluster (e.g. \citealt{1997MNRAS.292..419W}; \citealt{1998MNRAS.298..416P}; \citealt{2001MNRAS.322..589A}; \citealt{2010A&A...513A..37H}) would be of the order of $\sim 10^{2}-10^{3}~\text{M}_\odot~\text{yr}^{-1}$, thus implying a massive stream of gas falling onto the central BCG (on the order of $10^{11} - 10^{12}~\text{M}_\odot$ within the cluster dynamical time).
Nonetheless, the search for this cold ICM found far less cold gas and young stars than predicted (e.g. \citealt{1987MNRAS.224...75J}; 
\citealt{2005MNRAS.358..765H}; 
\citealt{2012ApJS..199...23H}; \citealt{2016A&A...595A.123M}).
Specifically, these works showed that local cool core clusters lack signatures of the massive cooling flows predicted by classical models since their central galaxies form new stars at $\sim 1$ \% of the predicted rate.
This fact became known as the \textit{cooling flow problem}.
A first answer to the problem came with the advent of the Chandra Observatory high-resolution X-ray imaging showing that the ICM in cool core clusters was not at all relaxed as supposed, but highly dynamic, due to the effects of BCG-hosted radio-loud AGNs (e.g. \citealt{2009ApJ...704.1586S}).
Indeed, the AGNs strong interaction with the ICM can be at the origin of large bubbles in the gas, possibly inflated by the radio jets (e.g. \citealt{2004ApJ...607..800B}; 
\citealt{2007ApJ...665.1057F}; 
\citealt{2015ApJ...805...35H}) and rising in size buoyantly, as well as outflows, cocoon shocks, sonic ripples and turbulent mixing (e.g. \citealt{2013MNRAS.432.3401G}).
The amount of mechanical energy released by AGNs is thought to counterbalance the ICM radiative losses due to the cooling effectively (e.g. \citealt{2007ARA&A..45..117M}; \citealt{2012ARA&A..50..455F}; \citealt{2012NJPh...14e5023M}), thus suggesting that `mechanical feedback', heating the ICM, could prevent the massive inflows of gas predicted by classical models with BCGs accreting inefficiently via `maintenance-mode' (e.g. \citealt{2005MNRAS.359..781S}, also referred as `low-excitation' or `radio-mode').
Spatially-resolved X-ray spectroscopic observations carried on with XMM-Newton and Chandra in the early 2000s (e.g. \citealt{2001ASPC..234..351K}; \citealt{2001ApJ...554..981P}; \citealt{2003ApJ...590..207P}; \citealt{2006PhR...427....1P}) revealed that the bulk of the ICM cooling was being quenched at temperature above one third of its virial value ($\sim 1$~keV).
These spectroscopic observations, corroborating findings from the \textit{FUSE} satellite and the Cosmic Origins Spectrograph on \textit{HST}, set upper limits on the amount of cooling material below $\sim 10^7$ K, lowering previous estimates of an order of magnitude ($\sim 10$\%), i.e. $\sim 30\ \text{M}_\odot\ \text{yr}^{-1}$ (e.g. 
\citealt{2003A&A...412..657S}; \citealt{2005ApJ...635.1031B}; 
\citealt{2014ApJ...791L..30M}; \citealt{2017ApJ...835..216D}). 

As a consequence of all these findings, nowadays cooling flows are not believed anymore to explain the stellar mass growth of BCGs but their on-going star formation and AGN activity through thermally unstable cooling of the ICM into warm and cold clouds sinking towards the BCG (e.g. \textit{chaotic cold accretion}; \citealt{2017ApJ...837..149G}; \citealt{2018ApJ...865...13T}). 
Still elusive, the BCGs formation scenario seems to be driven by galactic cannibalism, i.e. the merging of the BCGs progenitors with satellite galaxies which gradually sink towards the cluster centre due to dynamical friction. 
The timescale of this process is inversely proportional to the satellite mass, hence promoting preferential accretion of massive cluster companions. 
Recent simulations seem to point out that after an initial monolithic build-up of the BCGs progenitors and the virialisation of the hosting halo, the mass growth of BCGs at $z\lesssim 1$ is dominated by galactic cannibalism, and specifically through \textit{dry} (or dissipationless) mergers, i.e. gas poor and involving negligible star formation (e.g. \citealt{2003ApJ...597L.117K}; \citealt{2007MNRAS.375....2D}; \citealt{2016ApJ...816...86V}).
In this scenario for the hierarchical formation of galaxies, star formation at $z \lesssim 1$ should contribute only marginally to the BCGs stellar mass as confirmed from studies on BCGs at low redshift (i.e. $z < 0.3$, e.g. \citealt{2009MNRAS.395..462P}; \citealt{2010ApJ...715..881D}; \citealt{2012MNRAS.423..422L}; \citealt{2014MNRAS.444L..63F}).

In this framework, we study the BCG inhabiting the cool core cluster Abell 2667, a richness class-3 cluster in the Abell catalogue \citep{1958ApJS....3..211A}, located in the southern celestial hemisphere (i.e.~$\text{RA(J2000.0)}=23^h51^m39.37^s$; $\text{Dec(J2000.0)}=-26^\circ05'02.7''$), at redshift $z=0.2343$.
Abell 2667 is known for its gravitational lensing properties, due to the presence of several multiple image systems and a bright giant arc in its core (e.g. \citealt{1998MNRAS.301..328R}, \citealt{2006A&A...456..409C}).
Because of its extremely high X-ray brightness (i.e. $L_X = (14.90\pm 0.56)\times 10^{44} h_{70}^{-2}$~erg~s$^{-1}$ within the $0.1-2.4$~keV band, \citealt{1998MNRAS.301..328R}), Abell 2667 is one of the most luminous X-ray galaxy clusters in the Universe.
According to the cluster's regular X-ray morphology \citep{1998MNRAS.301..328R} and the dynamics of its galaxies \citep{2006A&A...456..409C}, Abell 2667 has been classified as a dynamically relaxed cluster showing the additional characteristic of a drop in the ICM temperature profile within its central region.
Therefore, the cluster hosts a cool core with an estimated cooling time of $t_{cool} = 0.5$~Gyr, as derived from \textit{Chandra} data by \cite{2009ApJS..182...12C}.
Using the NRAO VLA Sky Survey radio data at 1.4 GHz, \cite{2015A&A...581A..23K} listed the Abell 2667 central galaxy as a radio-loud source with a radio power of P$= 3.16\times 10^{24}$~W~Hz$^{-1}$, thus corroborating the AGN classification of the galaxy by \cite{2013MNRAS.432..530R}.
The analysis of \textit{Chandra} data has confirmed the presence of an AGN in a recent work by \cite{2018ApJ...859...65Y}.
Specifically, the authors classify the BCG as a type 2 AGN (i.e. an AGN with its broad line region (BLR) obscured) with an unabsorbed rest-frame $2-10$ keV luminosity of $2.82^{+5.50}_{-1.70}\times10^{43}~\text{erg\ s}^{-1}$.
In addition to AGN activity, in \cite{2012ApJ...747...29R}, the authors reported the BCG is forming stars at a rate, inferred from the far-IR via the spectral energy distribution (SED) templates by \citealt{2009ApJ...692..556R}, equal to $8.7 \pm 0.2$~M$_\odot~\text{yr}^{-1}$.
Ultimately, the presence of narrow (FWHM$<1000$~\kms~in the source rest-frame; \citealt{2000A&AS..144..247C}) but strong hydrogen and nebular emission lines ([OII]$\lambda3727$, \hb, [OIII]$\lambda\lambda4959,5007$, \halpha) have been detected around the central galaxy.

Thanks to the last generation integral field unit spectrographs such as the Multi Unit Spectroscopic Explorer (MUSE), a new and wider window on the description of the physical processes characterising galaxies beyond the Local Universe has been opened.
Our work aims at exploiting the exquisite quality of these new data in order to investigate in more detail the nature of the Abell 2667 central galaxy, spatially resolving the phenomena that the galaxy is undergoing.

This paper is structured as follows.
After the description in Section~\ref{sec:data} of the data-sets used in this work, we present our methods and results in Section~\ref{sec:analysis_results}. 
Specifically, in Section~\ref{sec:galfit} we describe the BCG structural properties as retrieved from the analysis of \textit{Hubble} Space Telescope multi-wavelength imaging data while in Sections~\ref{subsec:stellar_component} and ~\ref{subsec:gas_component} we report the results from the MUSE analysis of the BCG stellar and gas components, respectively.
Ultimately, in Section~\ref{sec:conclusions}, we summarise our findings and report different plausible scenarios describing the observed phenomena taking place within the Abell 2667 BCG.

Throughout this paper, we adopt a Flat $\Lambda$CDM cosmology of $\Omega_M=0.307$, $\Omega_\Lambda=0.693$, H$_0=67.7$~\kms~Mpc$^{-1}$.
According to this cosmology, the age of the Universe at $z=0.2343$ is $10.92$~Gyr, while the angular scale is $3.844$~kpc~arcsec$^{-1}$.
If not differently stated, the star-formation rates and stellar masses reported in this paper are based on a \cite{2003PASP..115..763C} initial mass function (IMF) while the magnitudes are in the AB system.

\section{DATA}
\label{sec:data}

Abell 2667 was observed with the integral field spectrograph MUSE \citep{2014Msngr.157...13B} at the Very Large Telescope (UT4-Yepun) under the GTO program 094.A-0115(A) (P.I. J. Richard) on the 26th October 2014.
The MUSE pointing is composed of four exposures of $1800$s, each one centred on the BCG but with slightly different FoV rotations to mitigate systematic artefacts.
The raw-data are publicly available on the ESO Science Archive Facility\footnote{\url{http://archive.eso.org/cms.html}} and we reduce them through the standard calibrations provided by the \texttt{ESO-MUSE pipeline} \citep{2014ASPC..485..451W}, version 1.2.1.
The final data-cube has a field of view of 1 arcmin$^2$ (corresponding to $\sim 230~\textnormal{kpc}\times230~\textnormal{kpc}$), with a spatial sampling of $0.2$\arcsec~in the wavelength range $4750 - 9350$ \AA.
Throughout the observations, the source was at an average airmass of 1 with an average V-band (DIMM station) observed seeing of $\sim 0.96$\arcsec~(FWHM).
To reduce the sky residuals, we apply the \texttt{Zurich Atmosphere Purge} (\texttt{ZAP} version 1.0, \citealt{2016MNRAS.458.3210S}) using a \texttt{SExtractor} \citep{1996A&AS..117..393B} segmentation map to define sky regions.

In our work, in addition to MUSE, we make use of ancillary data from the \textit{Hubble} Space Telescope (\textit{HST}).
The publicly available and fully reduced (Hubble Legacy Archive\footnote{\url{https://hla.stsci.edu/}}) \textit{HST} observations of the Abell 2667 cluster were taken on the 9th-10th October 2001 under the GO program 8882 (Cycle 9; P.I. S. Allen).
The observations, covering a mosaic FoV wider than $1.8'\times1.8'$, were carried out with the WFPC2 in the \textit{HST} filters (WF3 aperture): F450W ($5\times 2400$s), F606W ($4\times 1000$s), F814W ($4\times 1000$s).

\section{ANALYSIS AND RESULTS}
\label{sec:analysis_results}
\begin{table*}
\centering
    \begin{tabular}{|l|c|c|c|c|c|c|c|c|c|c|c|}
	\hline
	\hline
	band & m$_{tot}$ & $\Delta$m$_{tot}$ & $n$ & ${\Delta}n$ & $R_e$ & ${\Delta}R_e$ & $R_{e,circ}$ & $b/a$ & ${\Delta}b/a$ & PA & ${\Delta}PA$\\ 
	  & [mag] & [mag] & & & [arcsec] & [arcsec] & [arcsec] & & & [deg] & [deg]\\
	\hline
    F814W   & $16.31$ & $16.21-16.50$ & $4.97$ & $4.44-5.52$ & $9.28$ & $6.83-11.01$ & $8.15$ & $0.77$ & $0.76-0.80$ & $50.10$ & $49.56-51.39$\\
    F606W   & $17.14$ & $17.01-17.32$ & $4.40$ & $3.86-4.72$ & $8.69$ & $6.81-10.42$ & $7.52$ & $0.75$ & $0.73-0.77$ & $49.80$ & $49.56-50.29$\\
    F450W   & $18.30$ & $18.18-18.39$ & $4.55$ & $4.33-4.72$ & $7.84$ & $6.72-9.19$  & $6.79$ & $0.75$ & $0.71-0.77$ & $46.78$ & $37.84-49.86$\\
    mean    & $ - $   & $ - $         & $4.64$ & $4.21-5.00$ & $8.60$ & $6.78-10.21$ & $7.49$ & $0.76$ & $0.73-0.78$ & $48.89$ & $45.09-50.51$\\
	\hline 
	\hline
	\end{tabular}
	\centering{\caption{List of the \textit{galfitm} output parameters (and their variation range) for the Abell 2667 BCG.}}
	\label{tab:galfitm_param}
\end{table*}

\begin{figure*}
    \centering
    \includegraphics[width=\textwidth, trim={0 2cm 0 0}, clip]{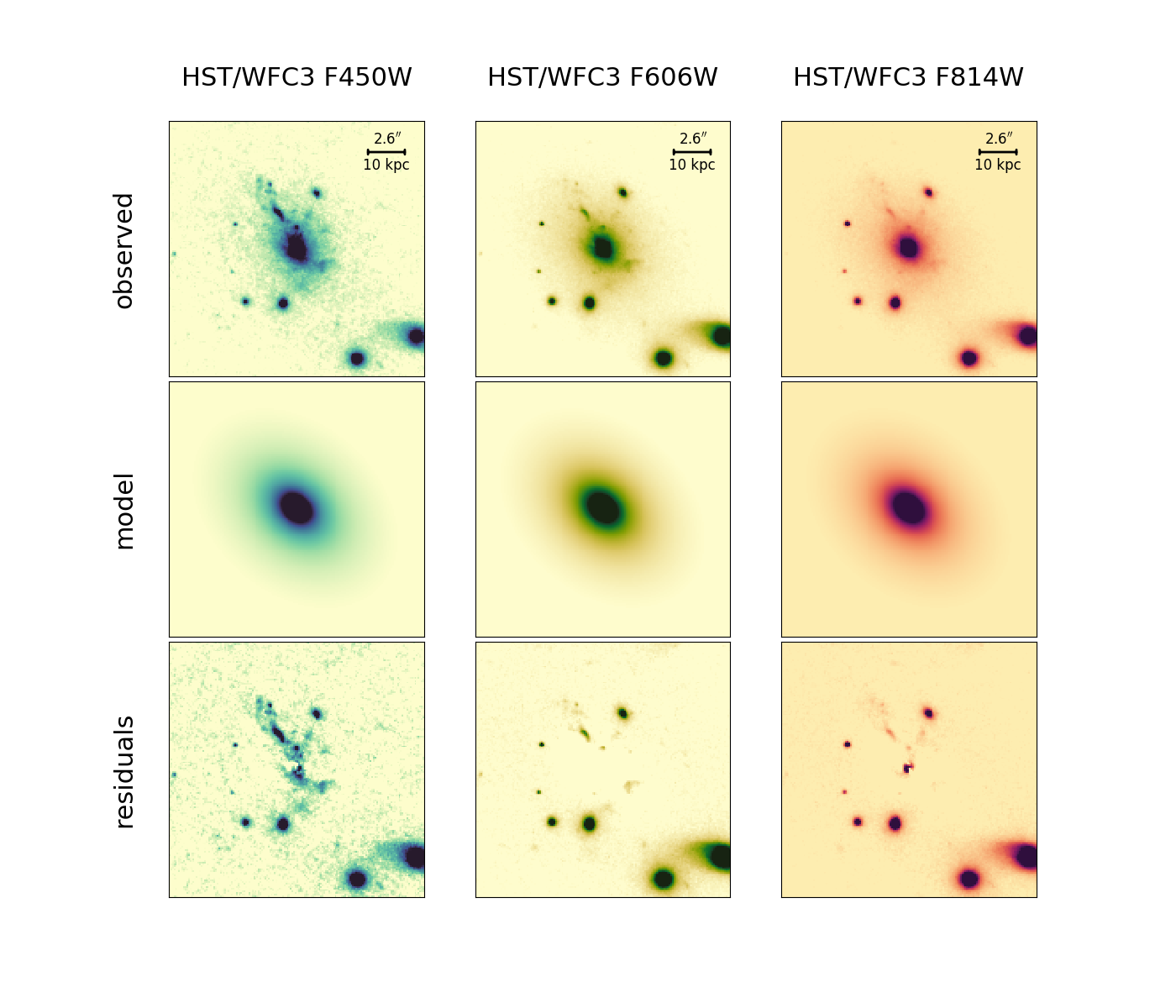}
    \caption{\textit{HST} images of the innermost region of Abell 2667 in the optical filters F450W, F606W and F814W (from left to right). From top to bottom, we present the original reduced \textit{HST} observations (P.I. S. Allen, \url{https://hla.stsci.edu/}), the two-dimensional \textit{galfitm} models of the BCG surface brightness profile and, ultimately, the map of the residuals, obtained subtracting from the original \textit{HST} images the \textit{galfitm} models.
    }
    \label{fig:rgb_galfit}
\end{figure*}

In the following section, we describe the analysis of both the \textit{HST} and MUSE data-sets which allow us to probe the BCG structural properties (e.g. total magnitude, S\'ersic\ index, effective semi-major axis, axis ratio, position angle), as well as its stellar and gas components.
The structural parameters are obtained thanks to \textit{galfitm} \citep{2013MNRAS.430..330H,2013MNRAS.435..623V} on \textit{HST} data.
From MUSE data we obtain results on the galaxy stellar component thanks to pPXF (i.e. penalised PiXel-Fitting, \citealt{2004PASP..116..138C}) and SINOPSIS (i.e. SImulatiNg OPtical Spectra wIth Stellar population models; \citealt{2007A&A...470..137F}), which allow for a study of the spatially resolved stellar kinematics, and the galaxy star formation rate (SFR), respectively.
Finally, the gas is studied employing tailored scripts.
Most of the results are achieved coding in Python and partially implementing the \texttt{Muse Python Data Analysis Framework} (\texttt{MPDAF}, \citealt{2016ascl.soft11003B}, \citealt{2016arXiv161205308C}, \citealt{2017arXiv171003554P}). 

\subsection{Structural Properties}
\label{sec:galfit}

The surface brightness profile of the BCG is fitted with a \cite{1968adga.book.....S} function in the F450W, F606W, and F814W \textit{HST} WFPC2 filters. 
To this aim, we use \textit{galfitm}, a multi-wavelength version of {\it GALFIT} \citep{2010AJ....139.2097P} that enables the simultaneous measurement of surface brightness profile parameters in different filters. 
The code allows the user to set the variation of each parameter (centroid coordinates, total magnitude m$_{tot}$, S\'ersic\ index $n$, effective semi-major axis $R_e$, axis ratio $b/a$, and position angle PA) as a function of wavelength, through a series of Chebyshev polynomials with a user-specified degree \citep[cf.][for further details]{2013MNRAS.435..623V}. 
We adopt a bright unsaturated and isolated field star as Point Spread Function (PSF) image.

One of the main problems to address when fitting the surface brightness profile of BCGs consists in accurately subtract the sky background, as well as the intracluster light (ICL) in the crowded central regions of the clusters.
In particular the ICL, if not considered, tends to increase both the size and the total luminosity of the BCG \citep{2007AJ....133.1741B}.
Hence, although estimating the light profile of the ICL is beyond the scope of this work, we took it into account by fitting the bright galaxies in the field, and the BCG+ICL, in separate steps, following an approach similar to that used in \cite{2016A&A...585A.160A}, as summarised below.
First, we fit together the BCG with all the (20) bright sources included in a field of view of $60''\times60''$ with single S\'ersic profiles, while masking the faintest objects, and the gravitational arcs present in the field.
In this first fit we left both the sky and the BCG S\'ersic parameters free as a function of wavelength, while forcing all the parameters (but the magnitudes) of the surrounding galaxies to remain constant in the three filters, to reduce the computational time (with Chebishev polynomial degrees set to 1 and 3, for the BCG and the other objects, respectively, \citealt{2013MNRAS.430..330H}).
In each filter, we derived the residual map of the BCG+ICL+sky by subtracting the best-fit models of the surrounding galaxies from the original image.
Then, we used this image, opportunely masked in the regions with strong residuals, to fit the BCG and the ICL separately.
We fitted single S\'ersic functions to both components, constraining the S\'ersic index of the ICL within the range $0.2-1$, since its surface brightness profile is known to have a flat core and sharply truncated wings (e.g., \citealt{2016A&A...585A.160A}, \citealt{2016A&A...590A..30M}).
In order to better estimate the sky background and the parameter uncertainties, we performed a series of \textit{galfitm} runs with different initial settings, by varying the Chebishev polynomial degrees.
We start (i) by forcing all the BCG and ICL parameters (but the magnitudes) to be constant as a function of wavelength in the fit.
Then, in the following runs, we increasingly relax the degrees of freedom of the system, by leaving as free parameters: ii) m$_{tot}$ and $R_e$; iii) m$_{tot}$, $R_e$, $n$; iv) all the parameters.
This is done for both the BCG and ICL, in all the allowed combinations, for a total of 16 different runs.
As expected, the best reduced $\chi^2$ is obtained for the fit with the largest number of degrees of freedom (all parameters free for both the BCG and ICL profiles). 
\footnote{In Fig.~\ref{fig:1d_surf_b_prof} (Appendix) we present the BCG surface brightness profile and its best-fit for the F814W filter.} 
However, we derive our estimates for the structural parameters of the BCG as the average among the results of all the \textit{galfitm} runs in each filter, as shown in Tab.~\ref{tab:galfitm_param}. 
In this table, instead of the formal \textit{galfitm} uncertainties (based on the $\chi^2$ test), which are unrealistically small, we report the variation range for each parameter. 
Mean values averaged-out among the three different filters are also shown.

From the results, we note that the derived circularised half-light radius of the BCG is in good agreement with the values reported in the literature (e.g. \citealt{2013ApJ...765...24N}, \citeyear{2013ApJ...765...25N}) while the highly centrally peaked S\'ersic profile ($n=4.5-5.0$) suggests that the galaxy is a massive dispersion-dominated spheroid.    

The two-dimensional models of the galaxy light profile in each \textit{HST} filter are then subtracted from the original observations, thus allowing for a visual estimate of the quality of the fit and, simultaneously, the detection of possible (sub-)structures not reproduced by the single S\'ersic profile. 
In Fig.~\ref{fig:rgb_galfit}, we present the zoomed-in cutouts ($\sim 16''\times 16''$) of the F450W, F606W and F814W \textit{HST} images before (upper panels) and after (bottom panels) the subtraction of the \textit{galfitm} model (middle panels).
In the top panel, there is clear evidence of clumpy structures together with a more diffuse emission extending along the spatial projected direction of the galaxy major axis, and in particular at the north-east and south-west extremities of the BCG.
The subtraction of the models from data highlights these clumps, showing they reside in filaments extending from the galaxy outskirts down to its very centre. 
Even though these substructures can be detected in all \textit{HST} bands, the most intense and structured emission comes from the bluer filter, i.e. F450W.
This fact suggests that the light coming from the filaments is principally originated by young stellar populations.

\subsection{Stellar Component}
\label{subsec:stellar_component}
In order to spatially resolve the properties of the BCG stellar component (stellar kinematics, mass building history) we resort to pPXF \citep{2004PASP..116..138C} and SINOPSIS \citep{2007A&A...470..137F}.
In the following subsections, we describe the results obtained from the two different codes, dividing the pPXF output (see Sec.~\ref{subsec:star_vel}) from SINOPSIS findings (see Sec.~\ref{subsec:star_pop}).

\subsubsection{Stellar Kinematics}
\begin{figure*} 
\centering
	\includegraphics[width=0.49\textwidth]{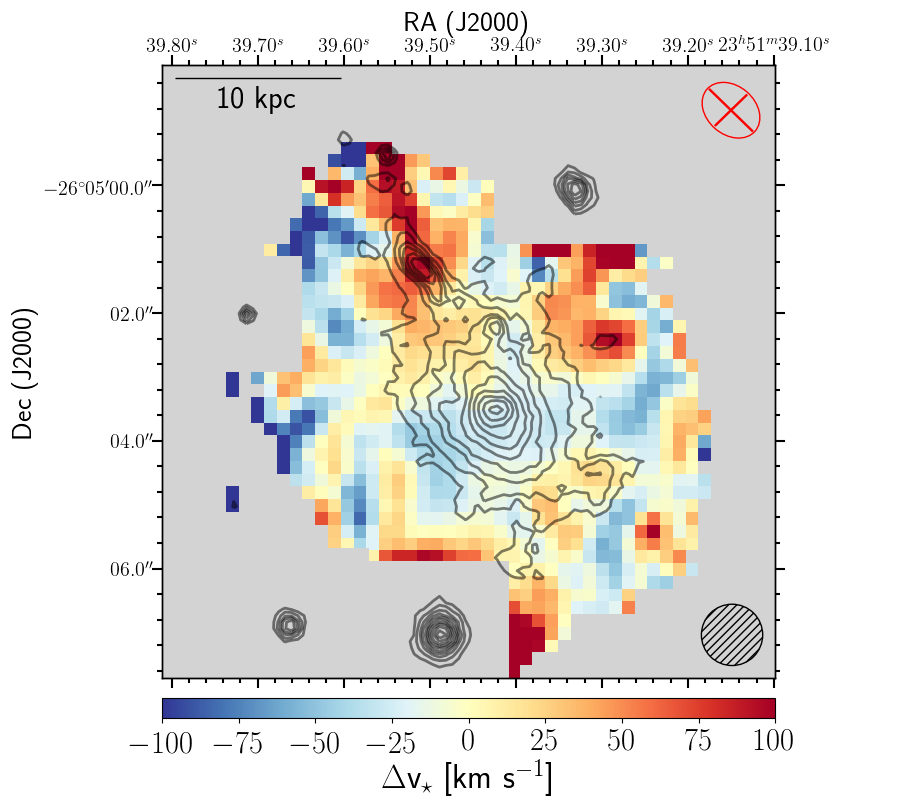}
	\includegraphics[width=0.49\textwidth]{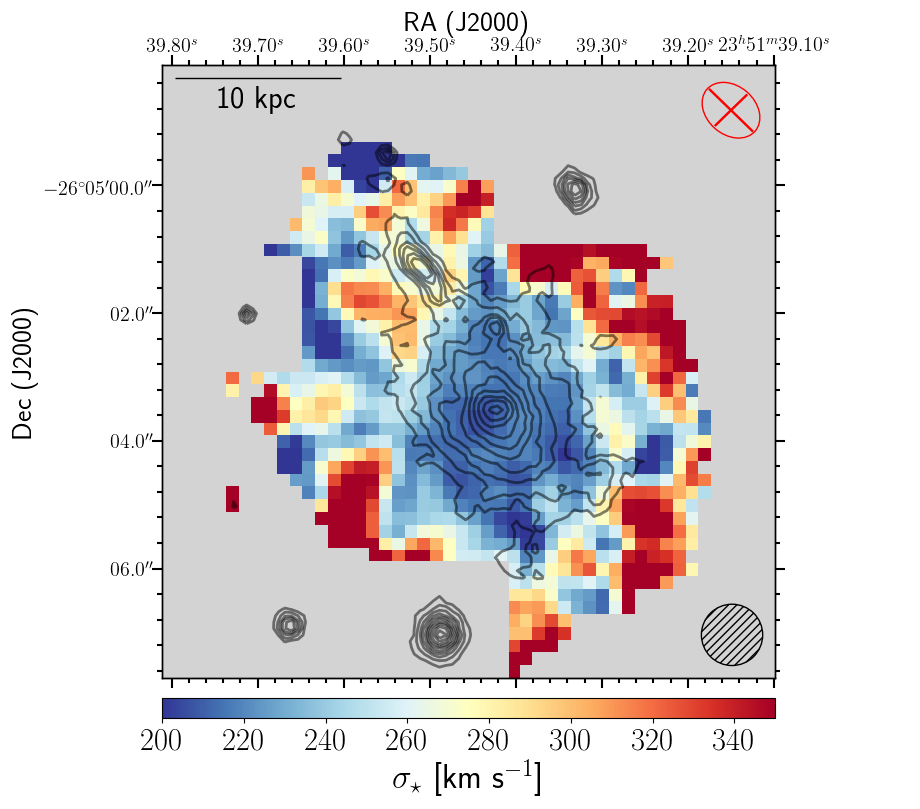}\\
	\caption{pPXF maps for the stellar velocity (\textit{left panel}) and velocity dispersion (\textit{right panel}) along the observer's line-of-sight, with the \textit{Hubble} Space Telescope F450W WFPC2 contours superimposed (\textit{grey solid lines}). In both maps, the spaxels with a SNR$<3$ are coloured in grey. The hatched black circle on the bottom right-hand side shows the beam size of the average V-band observed seeing of $\sim 0.96$\arcsec~(FWHM) while the red ellipse on the top indicates the galaxy orientation in the sky-plane. The ratio between the ellipse's axis corresponds to the $b/a$ value retrieved by the two dimensional surface brightness model of the galaxy by \textit{galfitm}.}
	\label{fig:stars_vel_sigma}
\end{figure*}
\label{subsec:star_vel}
\begin{table}
\centering
	\begin{tabular}{|l|c|c|}
	\hline
	\hline
	em.l. & $\lambda_\text{vacuum}$ [\AA] & e.m.l. type \\ 
	\hline 
	[NeIII] & 3868.760 & s \\
	HeI & 3888.648 & s \\
	H8 & 3889.064 & s \\
	$[\text{NeIII}]$ & 3967.470 & s \\
	H7 & 3970.075 & s \\
	$[\text{SII}]$ & 4068.600 & s \\
	H$\epsilon$ & 4076.349 & s \\
	H$\delta$ & 4101.760 & s \\
	H$\gamma$ & 4340.472 & s \\
	H$\beta$ & 4861.333 & s \\
	$[\text{OIII}]$ & 4958.911/5006.843 & db \\
	$[\text{NI}]$ & 5200.257 & s \\
	$[\text{OI}]$ & 6300.304/6363.776 & db \\
	$[\text{NII}]$ & 6548.050/6583.450 & db \\
	H$\alpha$ & 6562.819 & s \\
	$[\text{SII}]$ & 6716.440/6730.810 & db \\
	\hline 
	\hline
	\end{tabular}
	\caption{List of the most prominent emission lines (em. l.) detected in the Abell 2667 BCG spectra. The value of the rest-frame wavelength of each line (column 2) was taken from the NIST Atomic Spectra Database (\url{https://www.nist.gov/pml/atomic-spectra-database}). Finally, the third column reports whether the line is single (s) or a doublet (db).}
	\label{tab:emission_lines}
\end{table}
\begin{figure*}
\centering
	\includegraphics[width=\columnwidth,trim={0 2.85cm 0 0},clip]{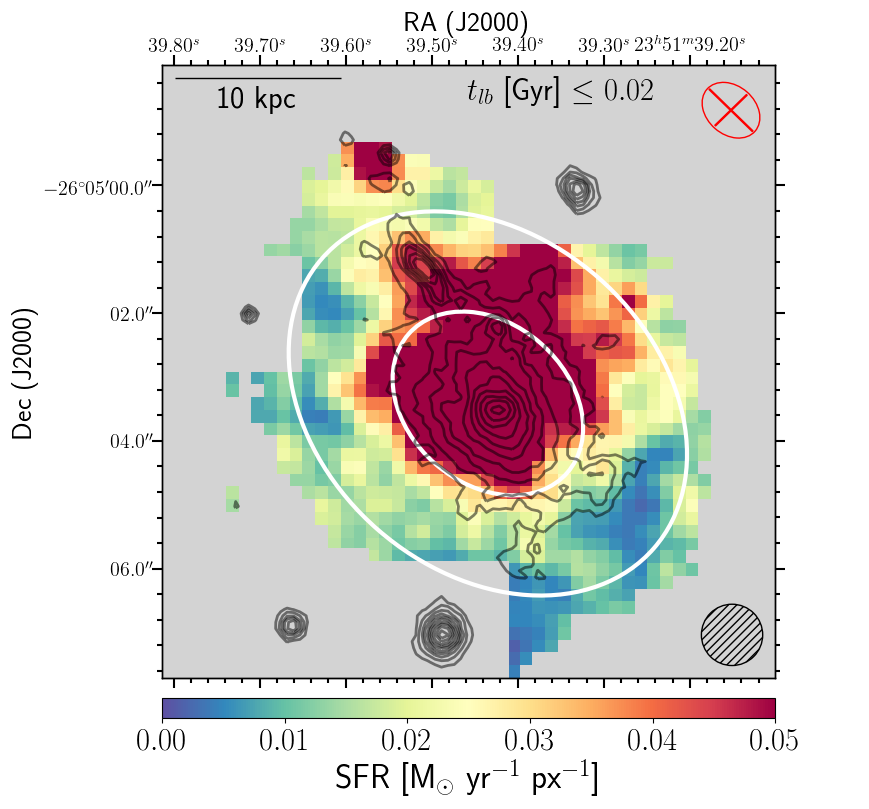}
	\includegraphics[width=\columnwidth,trim={0 2.85cm 0 0},clip]{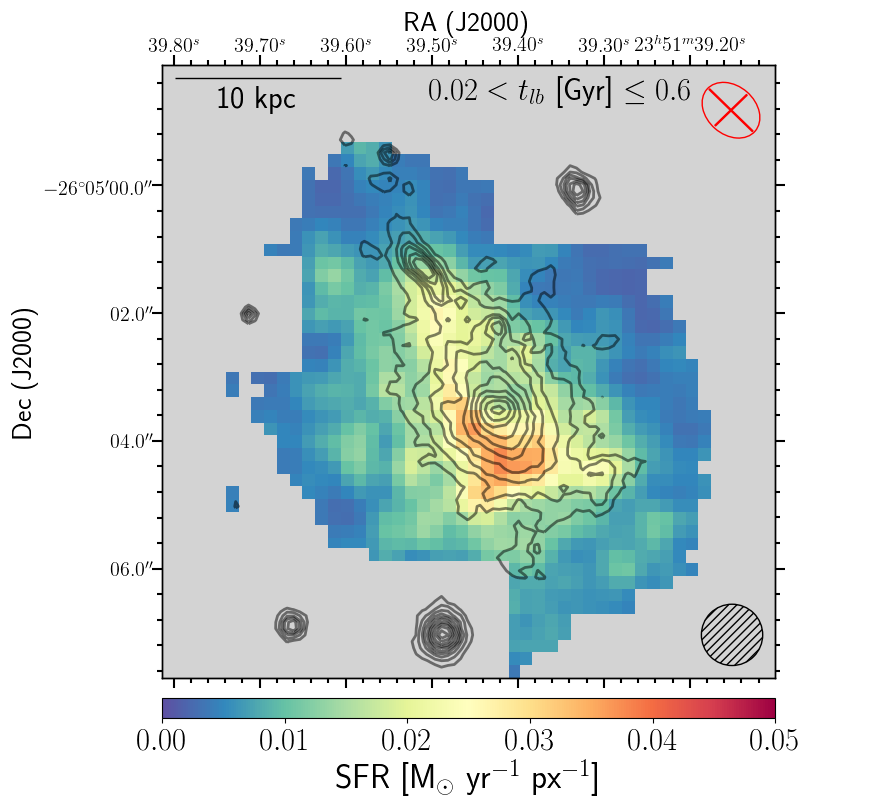}\\
	\includegraphics[width=\columnwidth,trim={0 0 0 0.95cm},clip]{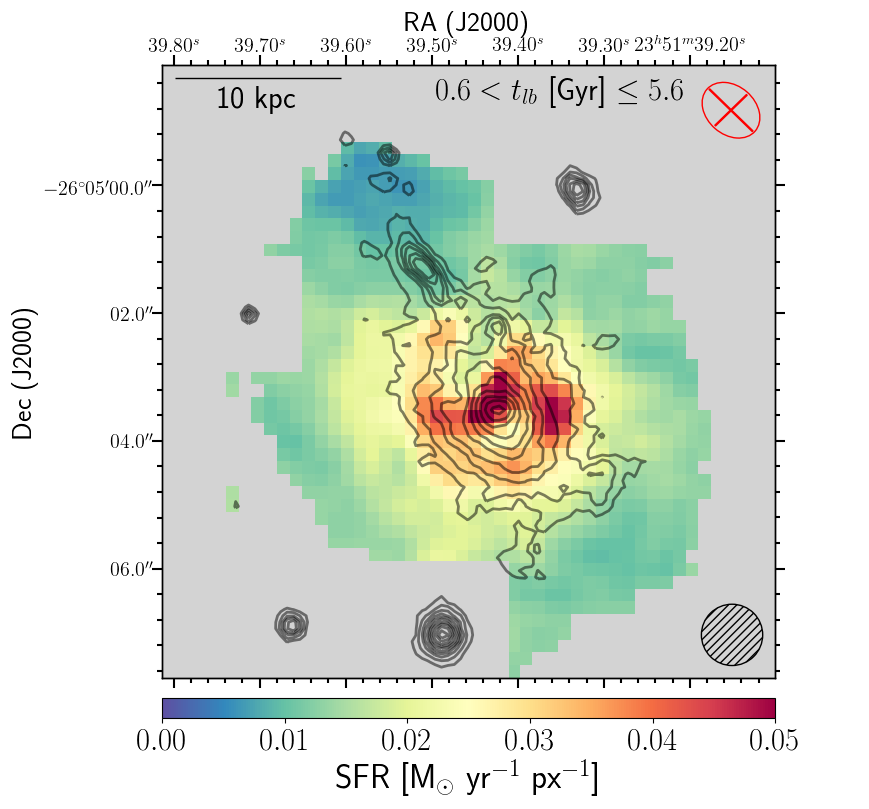}
	\includegraphics[width=\columnwidth,trim={0 0 0 0.95cm},clip]{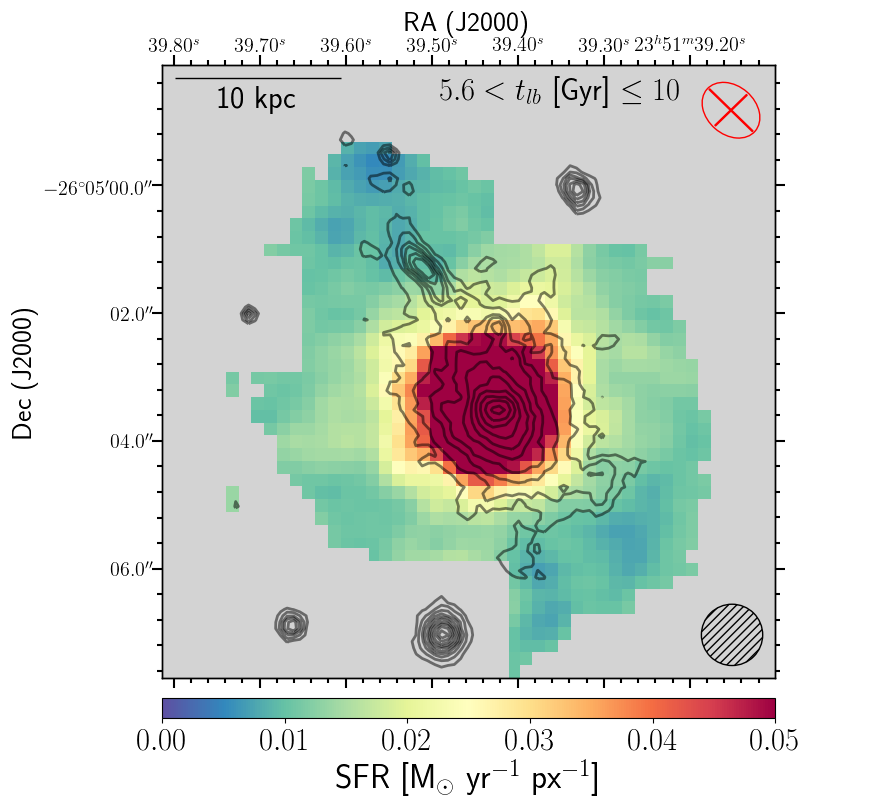}\\
	\caption{SINOPSIS SFR density maps of the Abell 2667 BCG in the look-back time ($t_{lb}$) intervals $t_{lb} \text{[Gyr]}\leq 0.02$ (\textit{top left}), $0.02 < t_{lb} \text{[Gyr]} \leq 0.6$ (\textit{top right}), $0.6 < t_{lb} \text{[Gyr]} \leq 5.6$ (\textit{bottom left}) and $5.6 <t_{lb} \text{[Gyr]} \leq 10$ (\textit{bottom right}), respectively. In the top left-hand side panel, the two white ellipses show the spatial extent of the regions we define as nuclear and outskirt in Fig.~\ref{fig:sfh}. Contours and symbols are defined as in Fig.~\ref{fig:stars_vel_sigma}.}
	\label{fig:stars_sfr}
\end{figure*}
\begin{figure*}
    \centering
    \includegraphics[width=\textwidth]{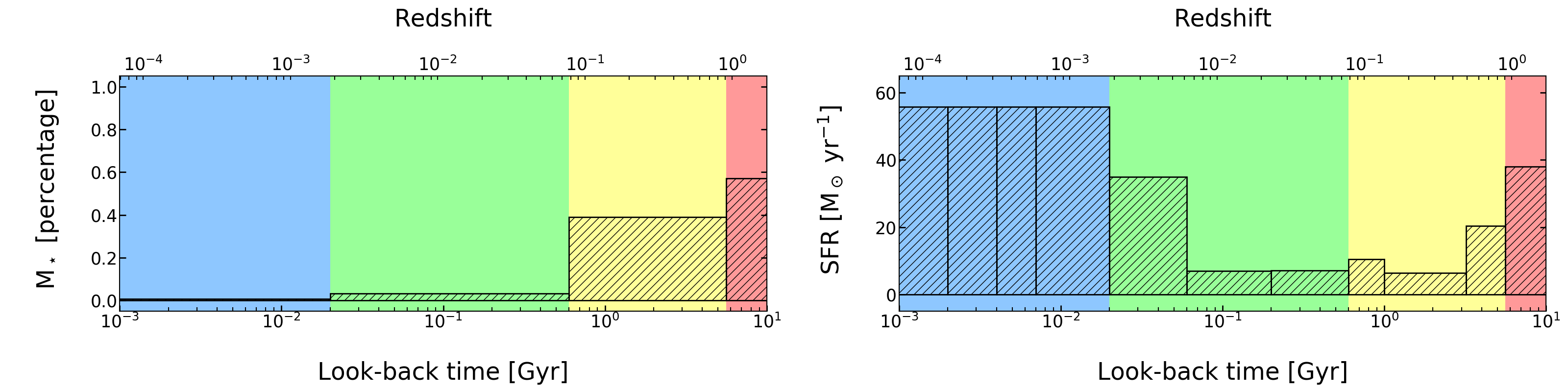}
    \includegraphics[width=\textwidth]{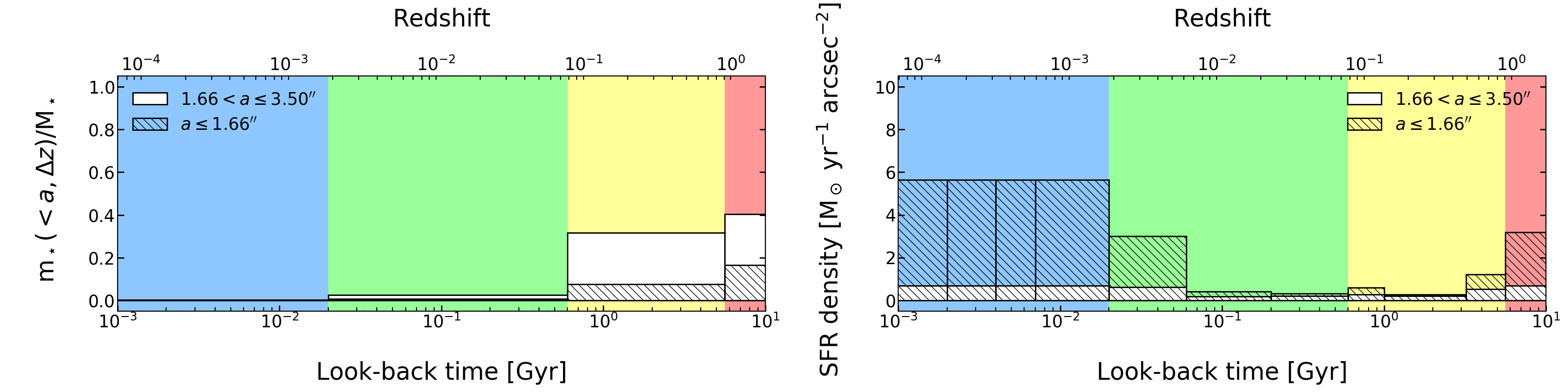}
    \caption{\textbf{Top panels}: SINOPSIS spatially integrated mass building (\textit{left panel}) and star formation history (\textit{right panel}) of the Abell 2667 BCG. The colour-shaded areas of the diagrams represent the four main age bins within which SINOPSIS calculates the galaxy percentage of mass built and SFR (see Fig.~\ref{fig:stars_sfr}). In the right panel, the histogram show the SFH computed in the highest temporal resolution mode (11 SSPs age intervals). Both the percentage of mass built and the SFR of each time-bin have been obtained integrating all the maps values within an elliptical regions centred at the galaxy coordinates,  with axis ratio $b/a = 0.76$, position angle PA$=48.89^\circ$ (for the parameters see Sec.~\ref{sec:galfit}) and semi-major axis $a=3.5''$.
   \textbf{Bottom panels}: Percentage of the stellar mass built (\textit{left panel}) and SFR density (\textit{right panel}) at different epochs for the nuclear (\textit{hatched histogram}) and external regions (\textit{white histogram}) of the BCG. For the nuclear region we have integrated SINOPSIS spatially-resolved values within an elliptical aperture centred at the galaxy coordinates, with $b/a$, PA as before, but $a=1.66''$. For the outskirts, we have resorted to an elliptical annulus with central coordinates, $b/a$ and PA as before, but with $1.66''<a\leq3.55''$. The two different areas are marked with the two white ellipses in the top left-hand side panel of Fig.~\ref{fig:stars_sfr}.}
    \label{fig:sfh}
\end{figure*}
To spatially resolve the BCG stellar kinematics we probe each spaxel (i.e. spatial pixels) of the MUSE data-cube where the signal-to-noise ratio (SNR) of the stellar continuum is $>3$. 
Specifically, we compute the SNR as the median value of the ratio between the observed flux and error in each channel of the data-cube corresponding to the $5250-6250$~\AA~rest-frame wavelength range. 
We adopt this interval of wavelengths since old stellar populations dominate its flux and there are no strong emission lines.
Whenever $\text{SNR} < 3$, we perform a 5th-order polynomial fit to the observed spectra in order to remove any deviation from linearity due to noise or bad sky reduction, whereas if $\text{SNR} > 3$, we proceed with a pPXF fit implementing the UV-extended ELODIE models by \cite{2011MNRAS.418.2785M}.
These models are based on the stellar library ELODIE \citep{2007astro.ph..3658P} merged with the theoretical spectral library UVBLUE \citep{2005ApJ...626..411R}.
The single stellar population models (SSPs) that constitute the UV-extended ELODIE models have a Salpeter initial mass function (IMF; \citealt{1955ApJ...121..161S}), a solar metallicity ($\text{Z} = 0.02$), a resolution of $0.55$~\AA~(FWHM) and a spectral sampling of $0.2$~\AA, covering the wavelength range $1000 - 6800$ \AA. 
The models span $54$ ages, ranging from $3$~Myr to $15$~Gyr.
To optimise the pPXF absorption-line fits, we mask the region about the most prominent emission lines (see Tab.~\ref{tab:emission_lines}) in the BCG spectra, along with the telluric lines at $5577$ \AA, $5889$ \AA, $6157$ \AA, $6300$ \AA~and $6363$ \AA, whose residuals could potentially contaminate the spectra.
We resolve to 6th-order additive polynomials and 1st-order multiplicative polynomials corrections for the continuum.

In Fig.~\ref{fig:stars_vel_sigma}, the maps of the stars line-of-sight velocity ($\Delta$v$_\star$) and velocity dispersion ($\sigma_\star$) are presented.
The stars line-of-sight velocities are relative to the BCG systemic velocity (i.e. v$_{sys}\simeq 1.36\times 10^5$~\kms) while the values of the velocity dispersion are corrected for instrumental broadening.
Although both maps show no evidence of a coherent kinematic pattern (e.g. rotation), stars along the same projected direction than the \textit{HST} blue filaments (see Sec.~\ref{sec:galfit}), appear to be all red-shifted.
This fact is indicative of their common spatial displacement, or motion, with respect to the BCG main stellar component. 

Our findings on the stellar kinematics suggest that the BCG is a dispersion supported system (as already pointed out by the value of the S\'ersic index in Sec.~\ref{sec:galfit}) because of the stars' low maximum velocity difference ($\Delta \text{v}_\star < 150$~\kms) and high central dispersion ($\sigma_0\sim 216\pm21$~\kms). 
The central dispersion of the galaxy is calculated averaging all the dispersion values in the right panel of Fig.~\ref{fig:stars_vel_sigma} within a circular aperture of 0.96\arcsec~(i.e. the seeing FWHM) centred at the galaxy coordinates.

The BCG central velocity dispersion allows us to obtain a rough estimate of the mass of the galaxy supermassive black hole (SMBH) thanks to the M$_{SMBH}-\sigma_0$ relation by \cite{2012MNRAS.426.2751Z}:
\begin{equation}
\text{M}_\text{SMBH} \sim 28\times 10^8\ \sigma_{200}^4 \ \ \ [\text{M}_\odot]
\end{equation} 
where $\sigma_{200}$ is the stellar velocity dispersion in units of $200$ \kms.
According to the relation above, valid for elliptical galaxies in a cluster, the BCG SMBH mass is equal to $3.8^{+1.7}_{-1.3}\times 10^9\ \text{M}_\odot$, with the uncertainties obtained propagating the errors on $\sigma_0$.
In agreement with Fig. 3 in \cite{2009ApJ...699L..43S}, high values for the SMBH mass are typical of low excitation galaxies such as LINERs, while they are statistically seldom in high excitation galaxies (e.g. Seyferts).  

\subsubsection{Stellar Population}
\label{subsec:star_pop}
To derive the stellar populations' properties of the BCG, we use SINOPSIS \citep{2007A&A...470..137F}. 
SINOPSIS is a code that allows the user to derive spatially-resolved stellar mass and star formation rate maps of galaxies at different cosmological epochs, through the combination of theoretical spectra of single stellar population models (SSPs).
We refer the reader to \cite{2017ApJ...848..132F} for a complete description of the code and on further details on the adopted models.

For each spaxel of the data-cube, SINOPSIS measures the galaxy stellar mass (M$_\star$), taking into account both stars which are in the nuclear-burning phase, and remnants (i.e. white dwarfs, neutron stars and stellar black holes).
Summing the value of all the spaxels within an elliptical region centred at the galaxy coordinates, with semi-major axis $a = 3.5$\arcsec~($\approx 13.5$~kpc), axis ratio $b/a = 0.76$ and PA$=48.89^\circ$ (for the parameters see Tab.~\ref{tab:galfitm_param}), the estimated stellar mass content is M$_\star \sim 1.38 \times 10^{11} \text{M}_\odot$.

The BCG building history is reconstructed through the composition of the  multiple SSPs.
Although SINOPSIS obtains the best-fit of spectra using up to 11 SSPs of different age\footnote{In its standard implementation, SINOPSIS uses up to 12 SSPs of different age to reproduce the observed spectrum of a galaxy. However, according to the redshift of our target (z=0.2343), the age bin corresponding to SSPs older than $11$ Gyr has been excluded.}, the relatively high temporal resolution provided by these SSPs does not allow to recover the SFH as a function of stellar age, because of an intrinsic degeneracy in the typical features of spectra of similar age and different dust attenuation (for further details, see \citealt{2007A&A...470..137F}).
To overcome this degeneracy, the 11 age bins are then re-arranged into four epochs that display the main stages of the galaxy building history.
The four epochs are carefully chosen to maximise the differences in the spectral features of the different stellar populations. 
For each one of the four age intervals, we present the BCG maps of SFR in Fig.~\ref{fig:stars_sfr} and the spatially-integrated SFR and percentage of mass built in Fig.~\ref{fig:sfh} (\textit{top panels}).
While the SFR for look-back time $>0.02$~Gyr is obtained from the best-fit relative weights of the different SSPs reproducing the spectra continuum, the SFR at epochs $\leq 0.02$~Gyr (i.e. the current SFR) is instead retrieved from the emission lines (see \citealt{2017ApJ...848..132F}).
Even though different mechanisms can be at the origin of the emission lines (e.g. star formation, AGN activity), SINOPSIS does not distinguish between these different processes and interprets the total flux of the emission lines as originated by only star formation.
Because our source is known to host an X-ray obscured AGN, the values of the current SFR have to be considered upper limits.
For this reason in Sec.~\ref{subsec:dust_extinction}, as a consequence of a more accurate fit of the emission lines in the spectra, we resort to spectroscopic diagnostic diagrams (e.g. \citealt{1981PASP...93....5B}) to determine the predominant mechanism originating the lines in each spaxel of the MUSE data-cube.
Nonetheless, according to the maps of Fig.~\ref{fig:stars_sfr}, the bulk of the BCG (i.e. the $57.2^{+11.6}_{-11.7}$\% of the galaxy total stellar mass) was assembled more than $5.6$ Gyr ago ($z>0.7$), while spatially limited star formation (SF) episodes seem to have taken place in the outer regions at later times, forming the $39.1^{+12.2}_{-12.7}$\% ($0.6<t_{lb}\leq5.6$~Gyr), $3.1^{+1.6}_{-1.9}$\% ($0.02<t_{lb}\leq 0.6$~Gyr) and $0.6^{+1.2}_{-0.5}$\% ($t_{lb}\leq0.02$~Gyr) of the BCG M$_\star$, respectively.
These secondary bursts of SF have highly irregular shapes that could be interpreted as the results of accretion of gas from an external source, e.g. a merger.
The map of the current star formation, tracing the SF in the last $0.02$ Gyr, is obtained from the emission lines. 
The ongoing SF is indicative of a vigorous episode of star formation (SFR $\approx 55\ \text{M}_\odot\ \text{yr}^{-1}$) spreading out across the whole BCG, extending from the innermost regions to the outskirts. 
The map seems to trace the isophotes of the \textit{HST} B-band image very effectively, corroborating the hypothesis that the observed blue filamentary structure around the galaxy (see Sec.~\ref{sec:galfit}) could be formed of a young stellar population.
In the bottom panels of Fig.~\ref{fig:sfh}, we present the look-back time profiles of the percentage of the total stellar mass built (i.e. $\text{m}_\star(<a,\Delta z)/\text{M}_\star$, \textit{left panel}) and SFR density (i.e. $\text{SFR}/\text{A}(<a)$, \textit{right panel}) in the BCG's nuclear and outskirt regions.
The left panel shows that the nuclear region and the outskirts had a similar mass assembly history, suggesting that the galaxy mass assembly occurred through similar processes (e.g. major merger).
The only difference between the histograms is the relative contribution to the total stellar mass (i.e. the normalisation) whose trend, however, is in line with expectations: the nuclear region contributes the most to the total stellar mass.
For what concerns the SFR density histogram, we see the inner regions having a more intense SFR than the outskirts.
In the last $0.06$~Gyr the SFR has increased significantly. 
In particular, in the outskirts the SFR density at $\leq 0.06$~Gyr is comparable to the initial SFR density (i.e. $t_{lb}\geq 5.6$~Gyr), hence considerably higher than for  $0.06<t_{lb}\text{[Gyr]}<5.6$.
The outskirts trend is also followed by the nuclear region, however here the current SFR appears to be a factor 5 higher.
As already pointed out at the beginning of this section, this current value of SFR density has to be considered an upper limit since contaminated by the AGN activity.

\subsection{Gas Component}
\label{subsec:gas_component}
The spaxel-by-spaxel subtraction of SINOPSIS synthetic stellar continua (see Sec.~\ref{subsec:stellar_component}, an example of SINOPSIS fit of the BCG stellar continuum is presented in Fig.~\ref{fig:gal_spec} of the Appendix) from MUSE observed spectra allows us to recover a pure emission lines data-cube.
From its visual inspection, we detect a wide variety of line morphologies (e.g. blue-wings, double-peaked lines, single lines with a broad component), a clear sign of the presence of different ongoing physical processes taking place within the BCG.
Though a robust characterisation of the emission lines properties is not easy to recover in these conditions, we develop a Python code performing for each spaxel simultaneous gaussian line fitting for \hb, [OIII]$\lambda5007$, [OI]$\lambda6300$, \halpha, [NII]$\lambda\lambda6548,6583$ and [SII]$\lambda\lambda6716,6731$ (referred as [NII] and [SII] doublets respectively, hereafter).
The code reproduces the emission lines using up to three gaussian components.
The free parameters of the code are the peak position of the gaussians (i.e. the centroid velocity), their standard deviation ($\sigma$) and amplitude.
To reduce the significant number of total free parameters, we assumed the lines peak position and $\sigma$ of each component coupled to the others, hence assuming all the components originate from the same region of the host galaxy.
While the lines wavelength separation is fixed based on theory, the amplitudes are free to vary. 
The only exception is made for the [NII] doublet, whose ratio is constrained by atomic theory (e.g. \citealt{2000MNRAS.312..813S}).

\begin{figure*}
    \centering
    \includegraphics[width=\textwidth]{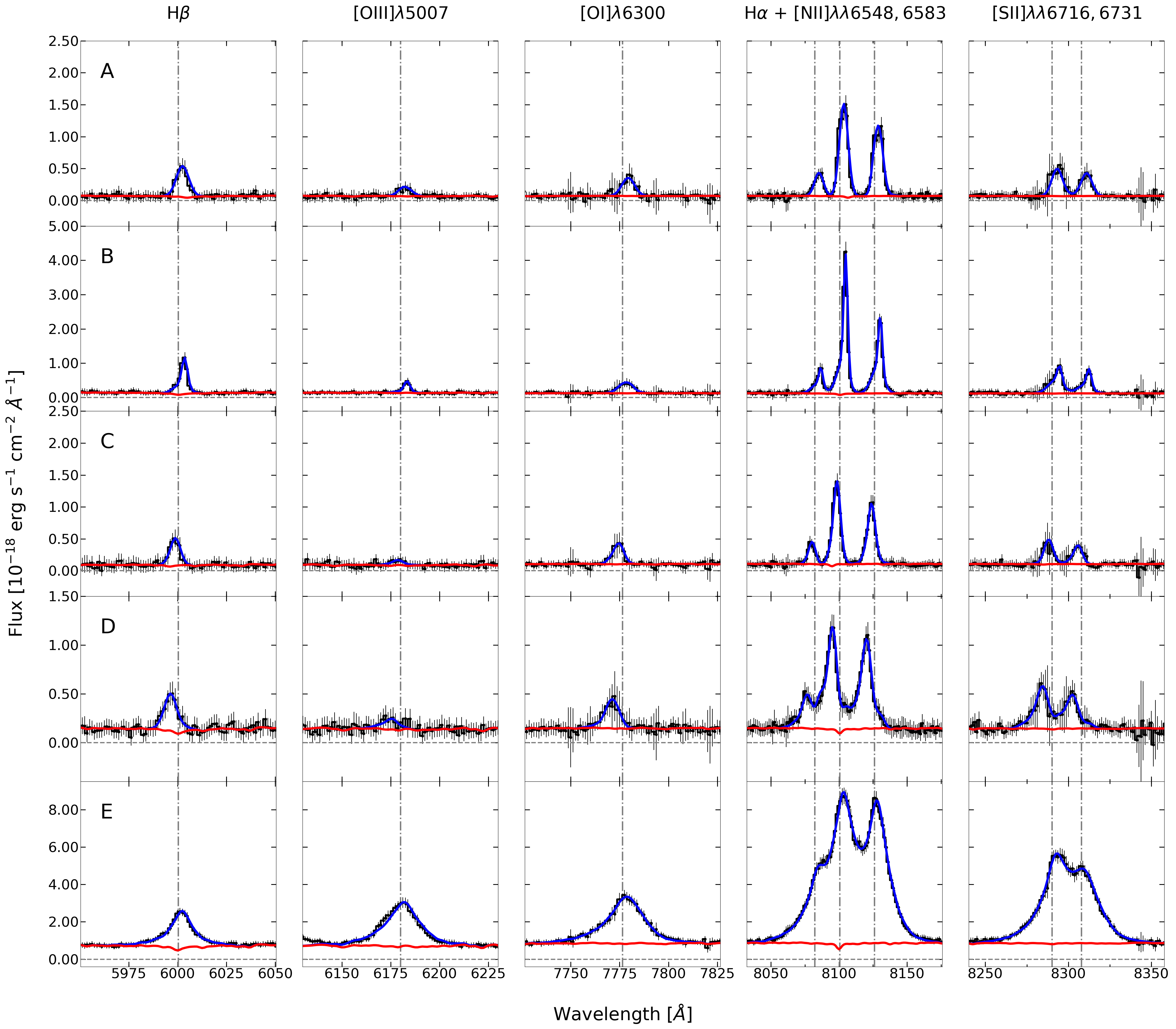}
    \caption{MUSE representative spectra and 3-$\sigma$ errors of the sky regions A, B, C, D, and E, presented in Sec.~\ref{subsec:velocity_channels}, Fig.~\ref{fig:hb_oiii_fluxes}. The \textit{red solid line} corresponds to the SINOPSIS model of the stellar continuum while the \textit{blue solid line} shows the fit of the spectra emission lines as obtained by our Python code (see Sec.~\ref{subsec:gas_component}). The vertical dash-dotted lines are indicative of the expected wavelength of each line, according to the BCG redshift.}
    \label{fig:fit_spectra}
\end{figure*}

We run the code for all the spectra with a clear detection (SNR $>3$) of the \halpha+[NII] doublet - i.e. the strongest and most spatially extended spectroscopic feature.
In Fig.~\ref{fig:fit_spectra}, we present the MUSE spectra, stellar-continuum and emission-line fits for five different spaxels of the BCG.
These spaxels are representative of very different regions of the galaxy and its surroundings, as the morphology of the lines suggests.
To estimate the errors on the parameters retrieved by our software, we perform for each spectrum 500 Monte Carlo simulations.
Each Monte Carlo run is achieved perturbing the stellar continuum-subtracted spectrum proportionally to its error, i.e. by varying the flux density according to a gaussian distribution centred on the observed flux and with $\sigma$ equal to the flux error. 
For each parameter, the standard deviation of the 500 Monte Carlo realisations is adopted as 1-sigma error.
The code outputs are the flux and associated error of each emission line, thus spatially resolving the emissions.
Ultimately, for each flux map, we discarded all the spaxels with an SNR lower than 3.

\subsubsection{Gas Kinematics}
\label{subsec:velocity_channels}
\begin{figure}
    \centering
    \includegraphics[width=\columnwidth]{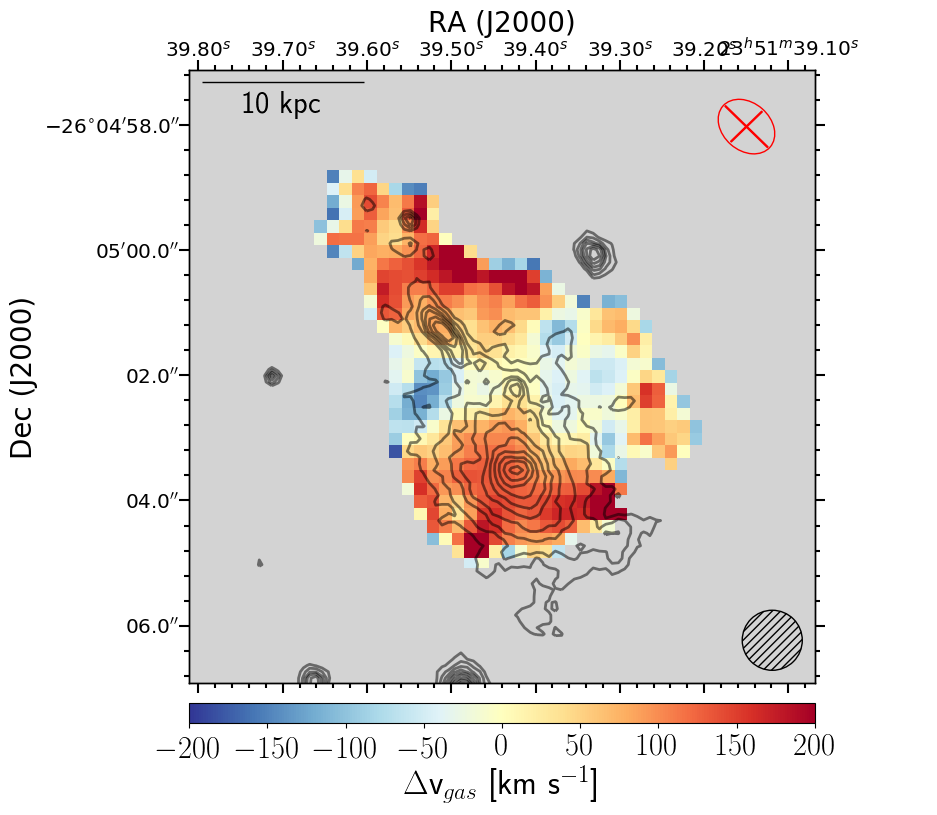}
    \caption{Map of the gas line-of-sight mean velocity ($\Delta\text{v}_{gas}$) with respect to the BCG rest-frame, with contours and symbols as defined in Fig.~\ref{fig:stars_vel_sigma}. The grey spaxels correspond to spectra with a SNR for the \hb $<5$.}
    \label{fig:gas_vel}
\end{figure}
Due to the variety in the shape of the emission lines that characterise the spectra in our data-cube, we calculate the gas line-of-sight velocity as the first moment of the distribution of the line fluxes.
To this aim, we choose to estimate the gas velocity from the \hb~since it is the strongest not-blended line in the spectra.
In Fig.~\ref{fig:gas_vel}, the map of the gas line-of-sight velocity with respect to the BCG rest-frame ($\Delta\text{v}_{gas}$) is presented.
The map is limited to the spaxels where the SNR of the \hb~is $> 5$. 
The $\Delta\text{v}_{gas}$ values obtained suggest that the overall gas emission is redshifted, from $\sim 50$~\kms up to $\sim 200$ \kms, spreading from the BCG nucleus all along the filamentary substructures revealed by \textit{HST}. 
Additionally, in small areas in between the main redshifted streams of gas, the gas velocity shows a negative trend  with average $\Delta\text{v}_{gas} \sim -100$~\kms, thus revealing the presence of a second blueshifted gas component.

\begin{figure*}
    \centering
    \includegraphics[height=.345\textwidth]{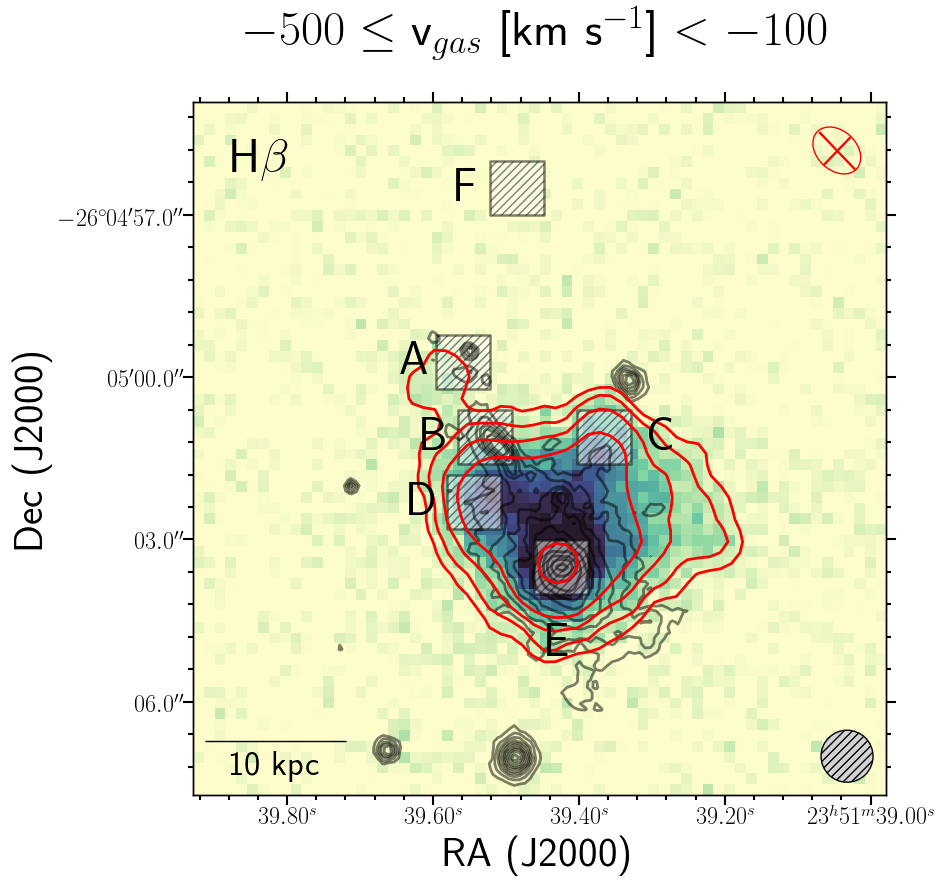}
    \includegraphics[height=.345\textwidth,trim={4.65cm 0 0 0},clip]{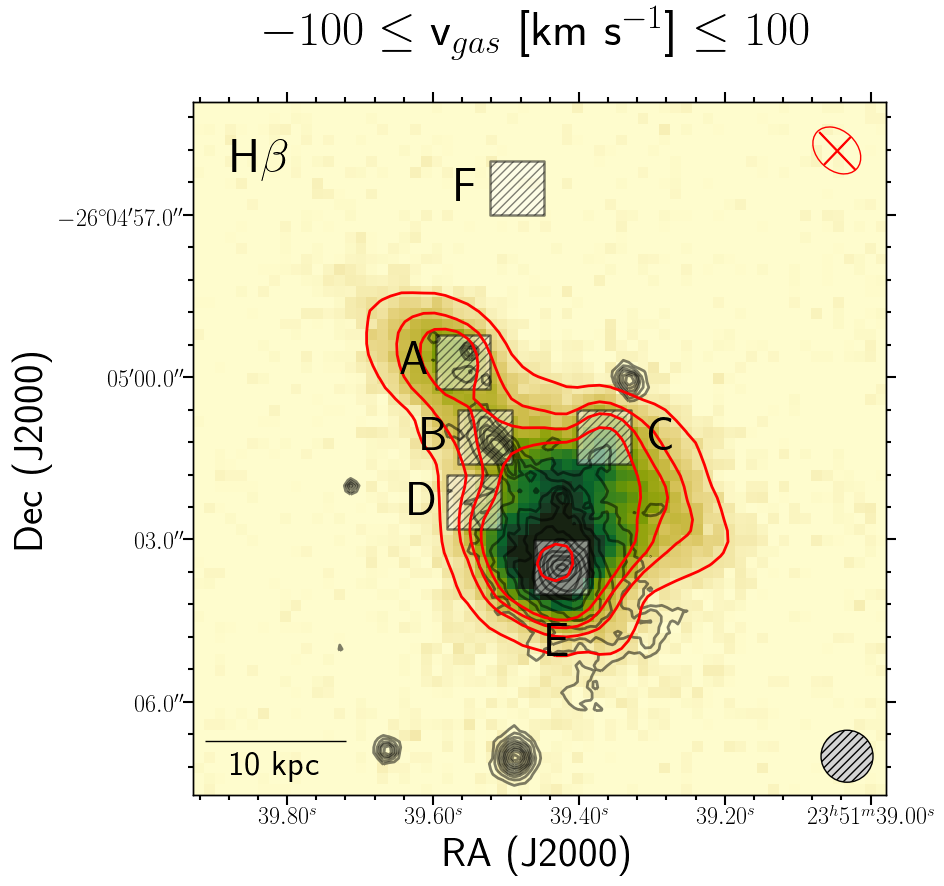}
    \includegraphics[height=.345\textwidth,trim={4.65cm 0 0 0},clip]{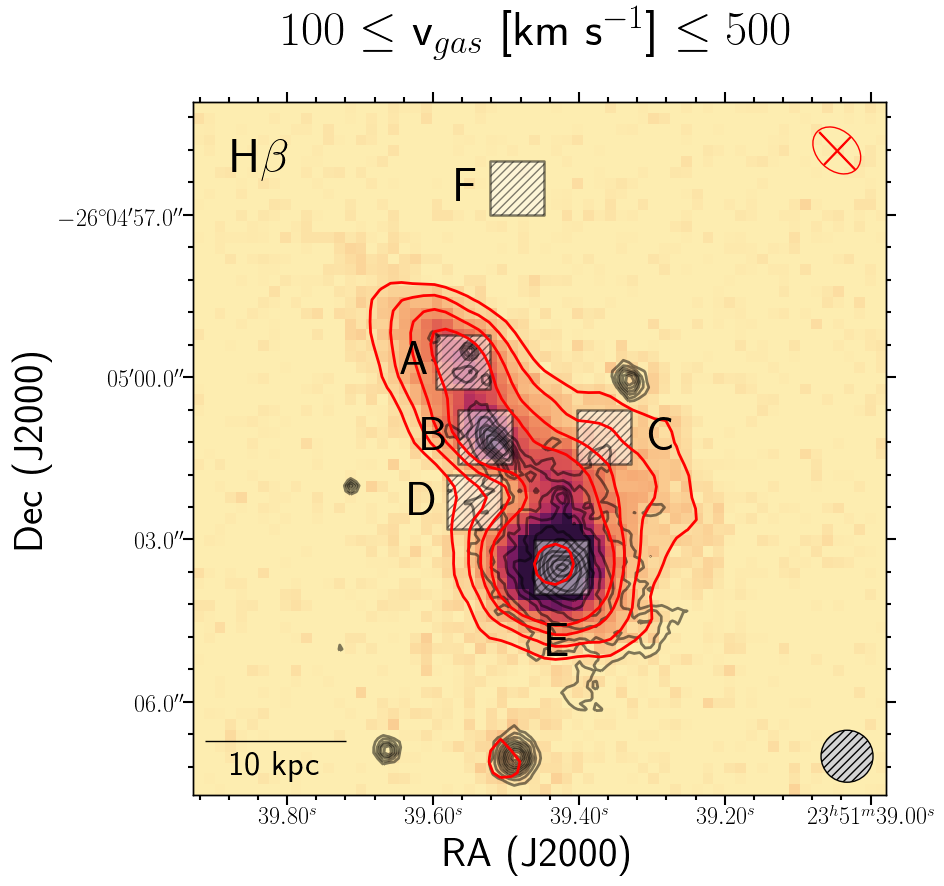}\\
    \includegraphics[width=\textwidth]{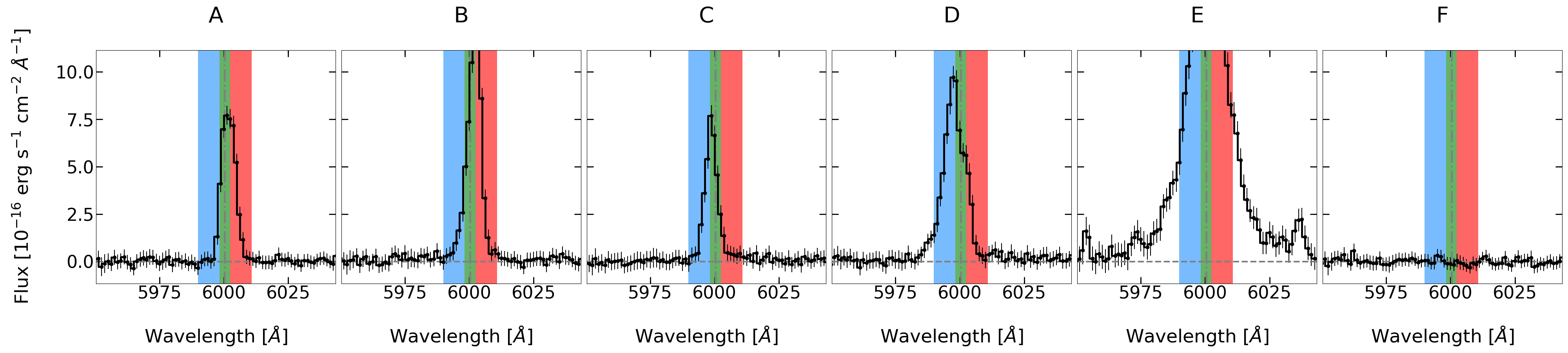}\\
    \includegraphics[height=.34\textwidth]{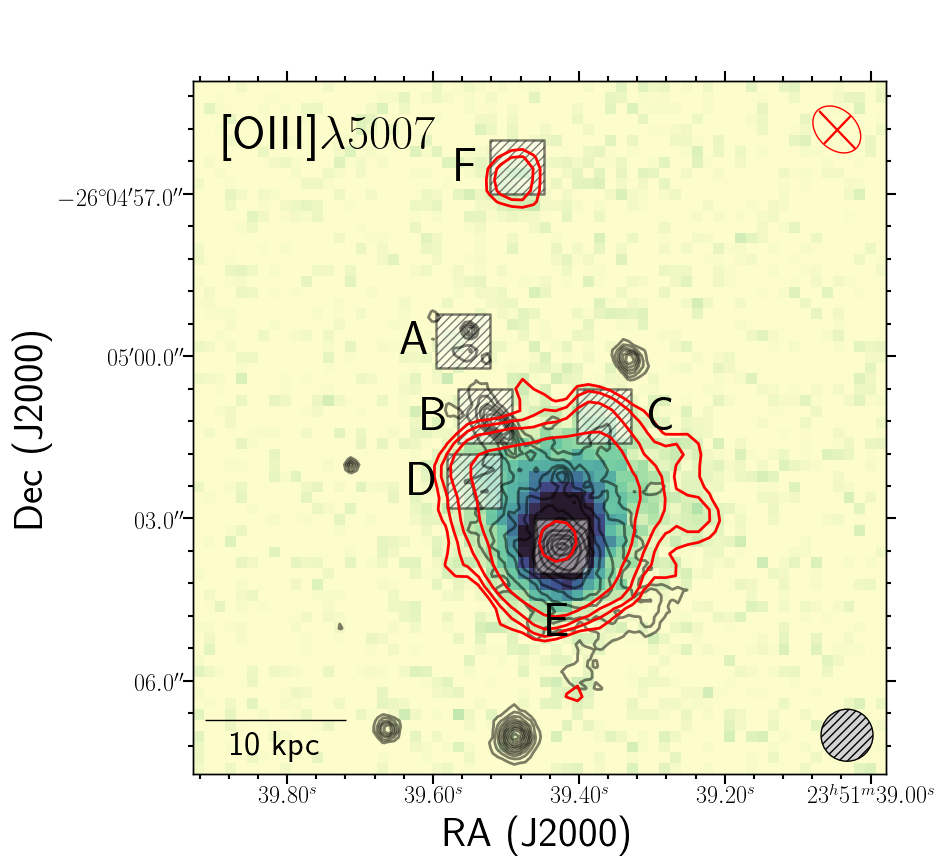}
    \includegraphics[height=.34\textwidth,trim={4.65cm 0 0 0},clip]{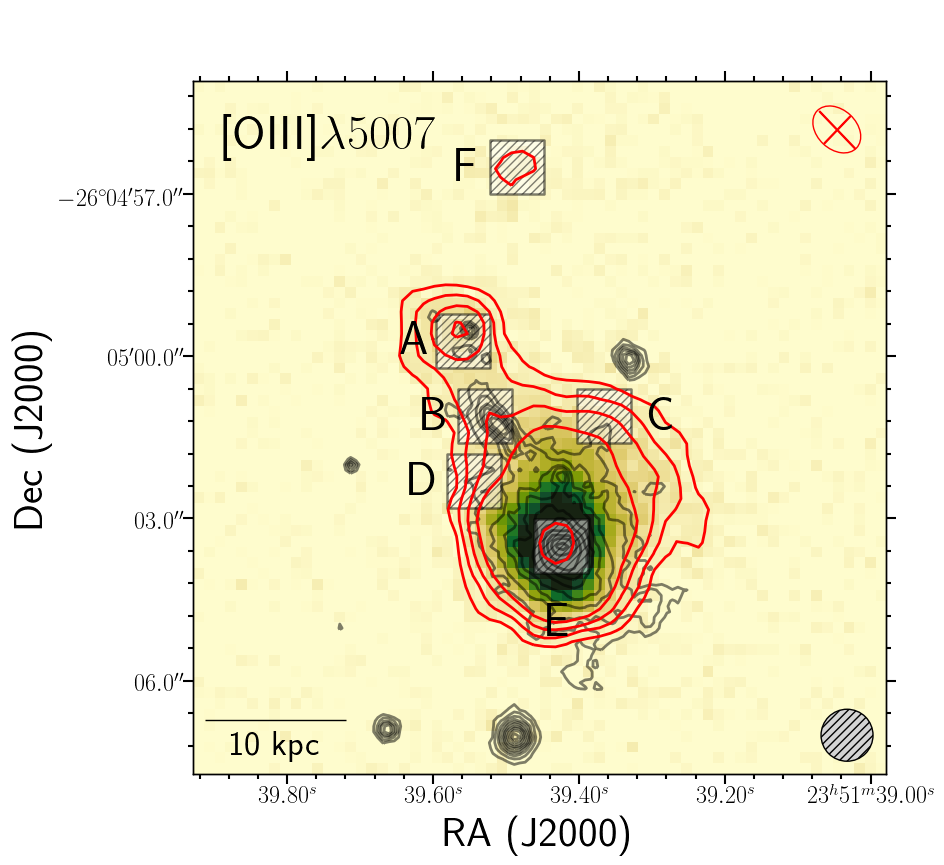}    
    \includegraphics[height=.34\textwidth,trim={4.65cm 0 0 0},clip]{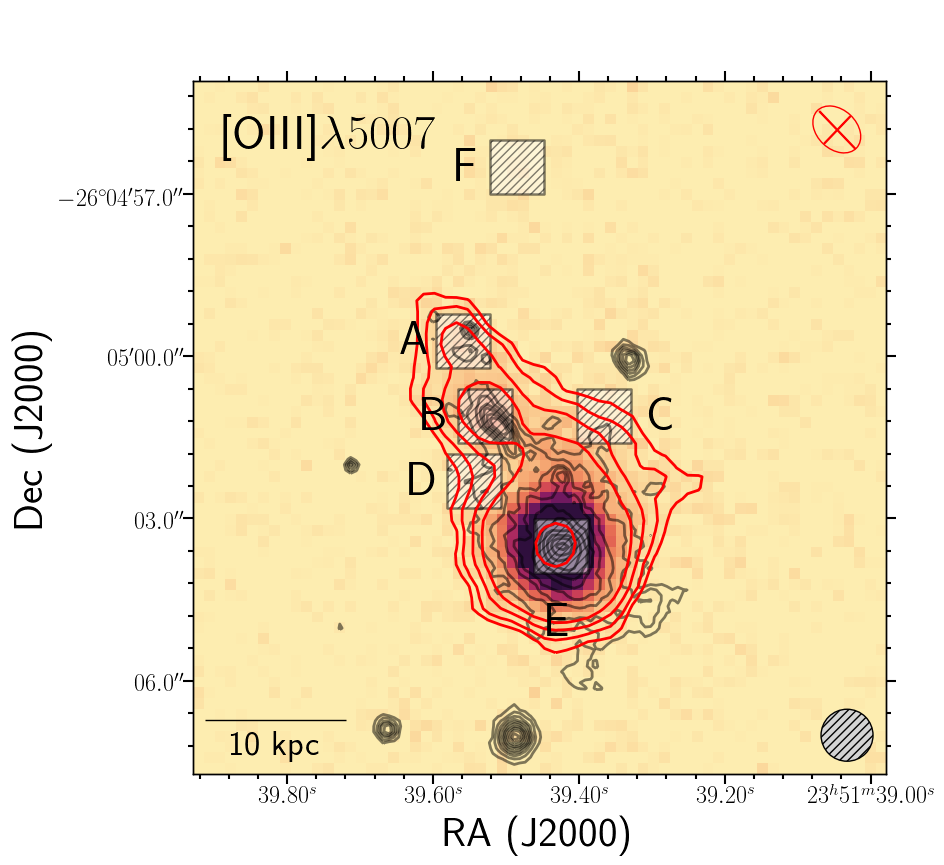}\\
    \includegraphics[width=\textwidth]{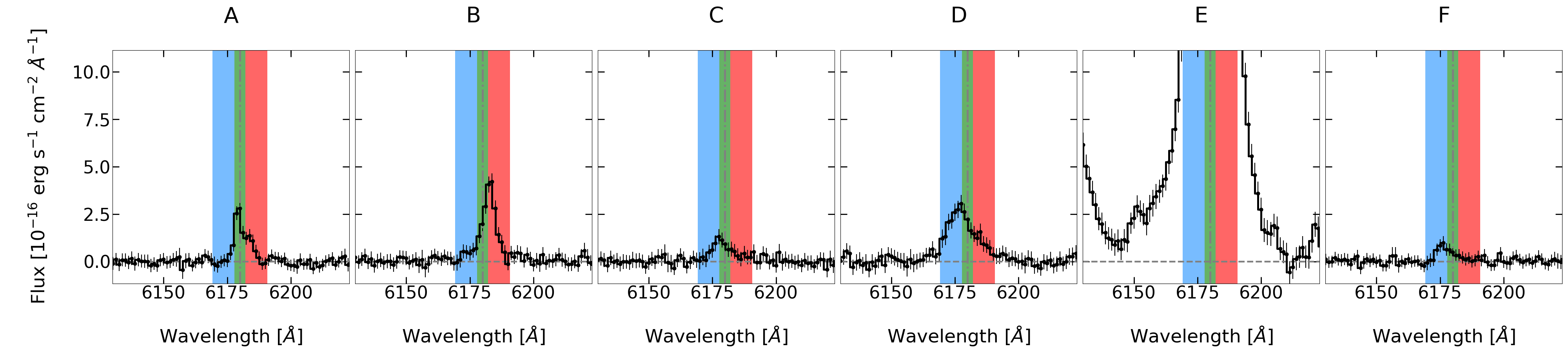}
    \caption{\hb~(\textit{top panels}) and \oiii~(\textit{bottom panels}) emission in Abell 2667 BCG with contours and symbols as defined in Fig.~\ref{fig:stars_vel_sigma}.. The images show the emission line integrated flux within three different velocity channels: from -500 to -100 \kms~(\textit{blue}), from -100 to 100 \kms~(\textit{green}) and from 100 to 500 \kms~(\textit{red}). In each panel, the \textit{red solid lines} are indicative of the $85.0$, $90.0$, $93.0$, $95.5$, $99.7$ percentiles of the emission lines flux, respectively. For both \hb~and \oiii, we report cutouts of the integrated spectra of six peculiar small (1\arcsec$\times$ 1\arcsec) sky regions, labelled with capital letters, and chosen to be representative of very different areas of the galaxy. In each cutout, the shaded areas indicate the different velocity ranges we adopted.}
    \label{fig:hb_oiii_fluxes}
\end{figure*}

To probe in greater detail the spatial extension of the emission lines and the gas velocity, we measure the integrated flux of both \hb~and \oiii~in different velocity channels (e.g. \citealt{2014ApJ...785...44M}), with respect to their rest-frame wavelength (see Tab.~\ref{tab:emission_lines}). 
The adopted bins cover the velocity ranges [-500;-100]~\kms, [-100;+100]~\kms~and [+100;+500]~\kms.
We inquire both the \hb~and the \oiii~lines because of the different physical processes to which these lines are particularly sensitive to, i.e. star formation and AGN activity.  
In Fig.~\ref{fig:hb_oiii_fluxes}, the integrated flux maps for \hb~and \oiii~are presented, highlighting in red the isocontours of the maps relative to the $85.0$, $90.0$, $93.0$, $95.5$, $99.7$ percentiles.
We also show the integrated spectra of six peculiar small (1\arcsec$\times$ 1\arcsec) sky regions and chosen to be representative of very different areas of the galaxy, as the variety in the emission lines shape suggests.
From the maps, we notice a difference in the projected spatial extent of the two emission lines, with the \hb~(\textit{top panels}) covering a wider area when compared to \oiii~(\textit{bottom panels}), which crowds just within the BCG and some of the knots of the \textit{HST} blue filamentary structure (see Sec.~\ref{sec:galfit}).
However, the main result of this analysis is the detection of two gas streams: an extended and red-shifted gas component- i.e. $\Delta$v$_{gas} = [+100;+500]$ \kms~- and a blue-shifted one - i.e. $\Delta$v$_{gas} = [-500;-100]$ \kms.
Comparing the projected spatial extension of these two gas streams with the isophotes from the \textit{HST} F450W image (see Fig.~\ref{fig:hb_oiii_fluxes}), the red-shifted gas, similarly to the stars velocity (see Sec.~\ref{subsec:star_vel}, Fig.~\ref{fig:stars_vel_sigma}), seems to trace the blue filamentary structure while the blue-shifted stream has a more roundish spatial extent.
Similarly to recent works in the literature (e.g. \citealt{2015A&A...582A..63C}, \citealt{2018A&A...619A..74V}), this blue-shifted gas component could be interpreted as a gas outflow originated by the activity of the central AGN.

From the maps of the \oiii, we observe at about $6''$ north of the BCG a weak feature (i.e. a blue spot) that has no counterparts for any of the other nebular lines in Tab.~\ref{tab:emission_lines} are detected.
Looking at the full integrated spectrum of the region (see Fig.~\ref{fig:lya_spec}) as well as the peculiar shape of the emission line (see spectrum F in Fig.~\ref{fig:hb_oiii_fluxes}, bottom panel), we interpret the blue spot as a serendipitous detection of an high-redshift line-emitting galaxy.
According to the observed wavelength of the emission, the assumed galaxy would be located at $z \simeq 4.08$, if the detected line is confirmed to be Ly$\alpha$. 

\subsubsection{Spectroscopic Diagnostic Diagrams}
\label{subsec:dust_extinction}
As already mentioned at the end of Sec.~\ref{subsec:star_pop}, we resort to empirical spectroscopic diagnostic diagrams, i.e. the BPT diagram by \cite{1981PASP...93....5B} and the diagrams by \cite{1987ApJS...63..295V} (see also \citealt{1995ApJ...455..468D}), to discriminate the physical process originating the emission lines detected in each MUSE spectrum.
This check is fundamental to retrieve a reliable value for the BCG current SFR from the map obtained by SINOPSIS (see Fig.~\ref{fig:stars_sfr}, top left-hand side panel).
Based on the intensity ratio [OIII]$\lambda5007$/\hb~versus [NII]$\lambda6583$/\halpha, [OI]$\lambda6300$/\halpha~and [SII]$\lambda6716+6731$/\halpha, these empirical diagrams are commonly adopted in determining the predominant ionisation mechanism rising the lines.
Specifically, the relative strengths of these prominent emission lines give insights into the nebular conditions of a source thus helping separate star-forming galaxies from AGNs (LINERs and Seyferts).
Thanks to the close wavelength of the lines involved, the diagrams do not suffer significantly from reddening correction or flux calibration issues.
For the sake of simplicity, hereafter we will refer to these diagrams as BPT-[NII] (the original BPT diagram by \citealt{1981PASP...93....5B}), BPT-[OI] and  BPT-[SII] (the diagrams by \citealt{1987ApJS...63..295V}). 
To define the loci of the diagrams populated by star-forming galaxies and AGNs we resort to the empirical relations by \cite{2001ApJ...556..121K}.
In the BPT-[NII] we adopt the equation by \cite{2003MNRAS.346.1055K} to isolate the so-called `transient' objects (e.g. \citealt{2008ARA&A..46..475H}, and references therein).

\begin{landscape}
\begin{figure}
\begin{center}
\includegraphics[scale=.19]{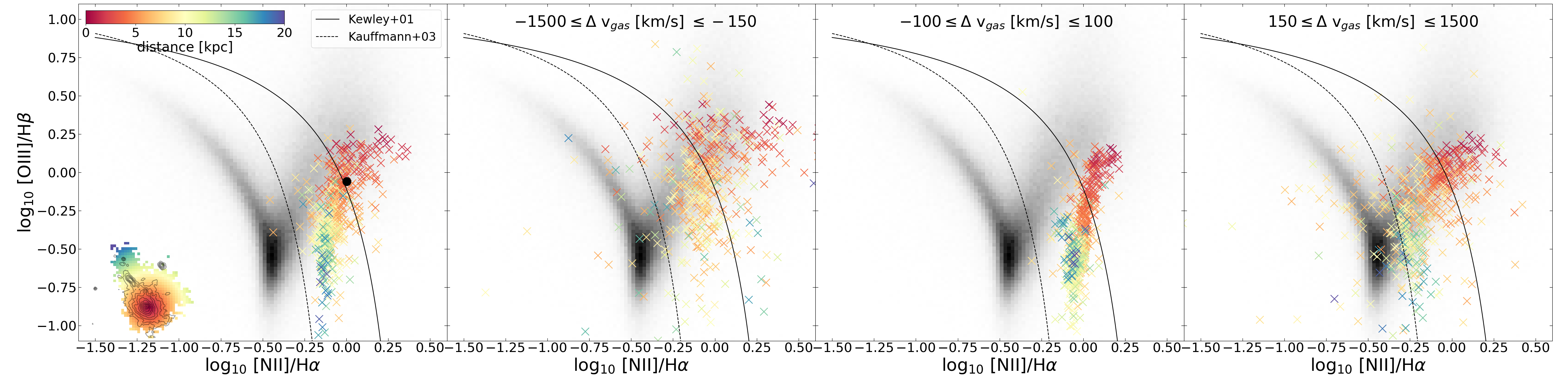}\\
\medskip
\includegraphics[scale=.19]{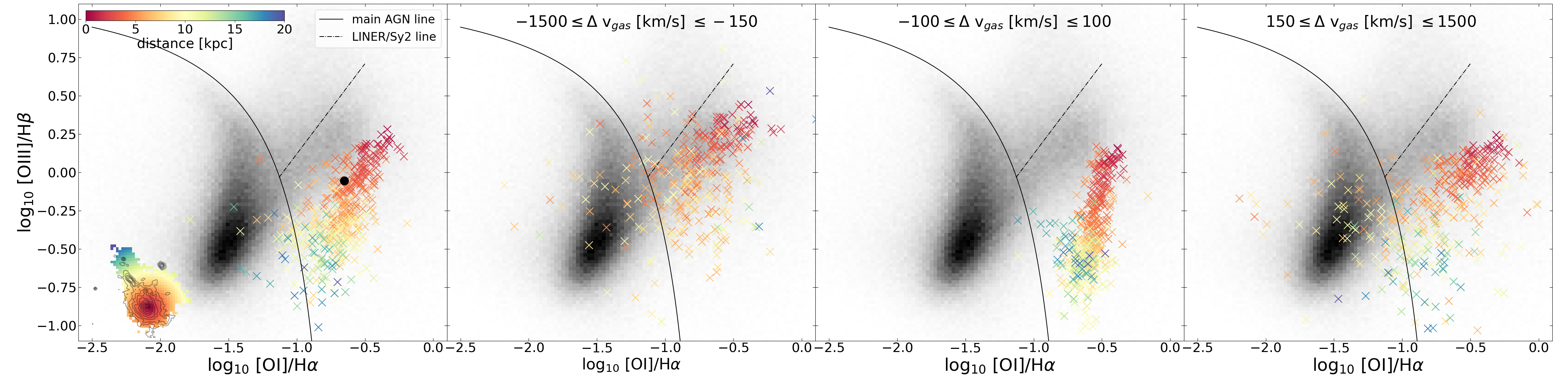}\\
\medskip
\includegraphics[scale=.19]{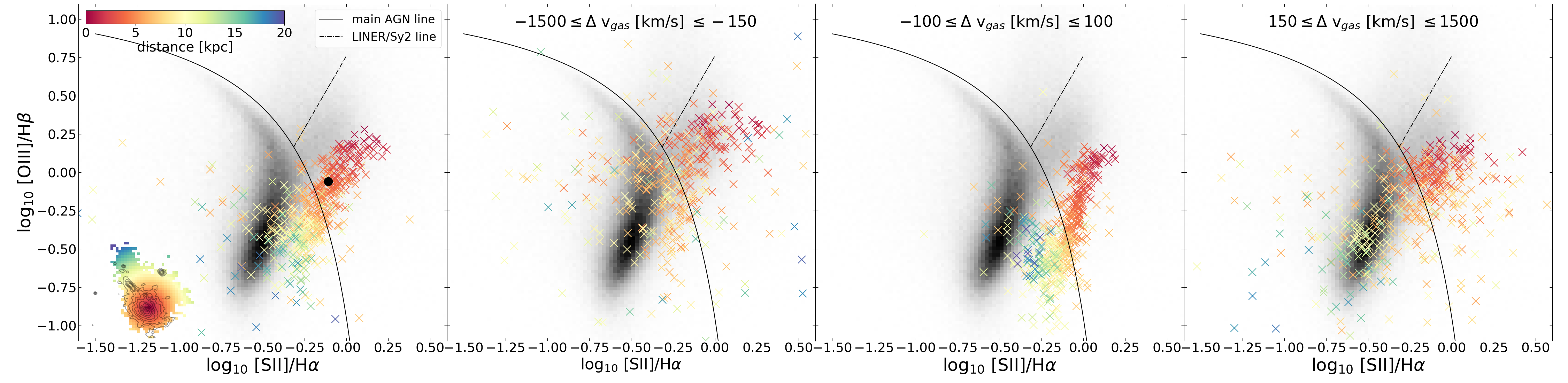}
\end{center}
\caption{From top to bottom, the BPT-[NII], BPT-[OI] and the BPT-[SII] diagrams for the spaxels with SNR([OIII]$\lambda5007) > 3$. For each diagnostic, from left to right we report the diagram obtained integrating all the fluxes within -1500 \kms~to 1500 \kms, and in the velocity ranges -1500 \kms~to -150 \kms (\textit{blueshifted outflow}), -100\kms~to 100 \kms (\textit{disc}) and 150\kms to 1500\kms (\textit{redshifted outflow}), respectively. Each mark is representative of a spaxel and is colour-coded according to the its distance from the BCG centre (see the colourbar on the top left and the colour map on the bottom left of the left side panels). To separate the loci corresponding to star-forming galaxies, transition objects and AGNs, we report the empirical relations by \citealt{2001ApJ...556..121K} and \citealt{2003MNRAS.346.1055K}. Finally, we report in black the distribution of the emission line galaxies in the Sloan Digital Sky Survey (SDSS) DR7 \citep{2009ApJS..182..543A}.}
\label{fig:bpts}
\end{figure}
\end{landscape}

Thanks to the multi-gaussian fitting procedure adopted (see Sec.~\ref{subsec:gas_component}), we can deblend the \halpha+[NII] emission as well as the [SII] doublet. 
From the deblended spectra, we reconstruct spatially resolved diagnostic diagrams, being able to resolve them also in velocity channels.
In particular, similarly to \cite{2018arXiv181107935M}, we define a `disc', and a blue and red-shifted `outflow' components with respect to the BCG rest-frame velocity.
While the `disc' is obtained from lines fluxes within -100 to 100 \kms, the blue and red-shifted components span the -1500 to -150 \kms~and 150 to 1500 \kms ranges, respectively.
All the diagnostic diagrams retrieved are presented in Fig.~\ref{fig:bpts} together with the empirical relations separating the different class of objects (\citealt{2001ApJ...556..121K}; \citealt{2003MNRAS.346.1055K}) and the distribution of the galaxies in the Sloan Digital Sky Survey (SDSS) DR7 \citep{2009ApJS..182..543A}.
The marks in the diagrams are representative of all the spaxel where the SNR of the [OIII]$\lambda5007 > 3$ and they are colour-coded depending on the distance of the spaxel from the BCG centre.
In each panel on the left, we additionally highlight with a black mark the position that the BCG occupies if we stack together all its MUSE spectra with SNR([OIII]$\lambda5007) > 3$.
The loci inhabited by the BCG in the diagrams define the galaxy unequivocally as a LINER.

If we spatially resolve the diagnostic diagrams, the line intensity ratios show that the very core of the BCG is consistent with an AGN, in agreement with \cite{2015A&A...581A..23K} and \cite{2018ApJ...859...65Y}, while the outer regions exhibit signs of both AGN and SF activity.
We hypothesise that these two physical mechanisms could be linked to the different components of the gas we reported in Sec.~\ref{subsec:velocity_channels}.
In particular, while the emission lines coming from the regions characterised by the red-shifted gas stream that is cospatial with the \textit{HST} blue filaments could trace the SF primarily, the blue-shifted gas component might be related to the AGN, thus tracing a possible feedback (i.e. outflow) from the BCG active galactic nucleus (e.g. \citealt{2015A&A...582A..63C}).

\subsubsection{Electron Density}
\label{subsec:electron_density}
\begin{figure*}
    \includegraphics[width=\columnwidth]{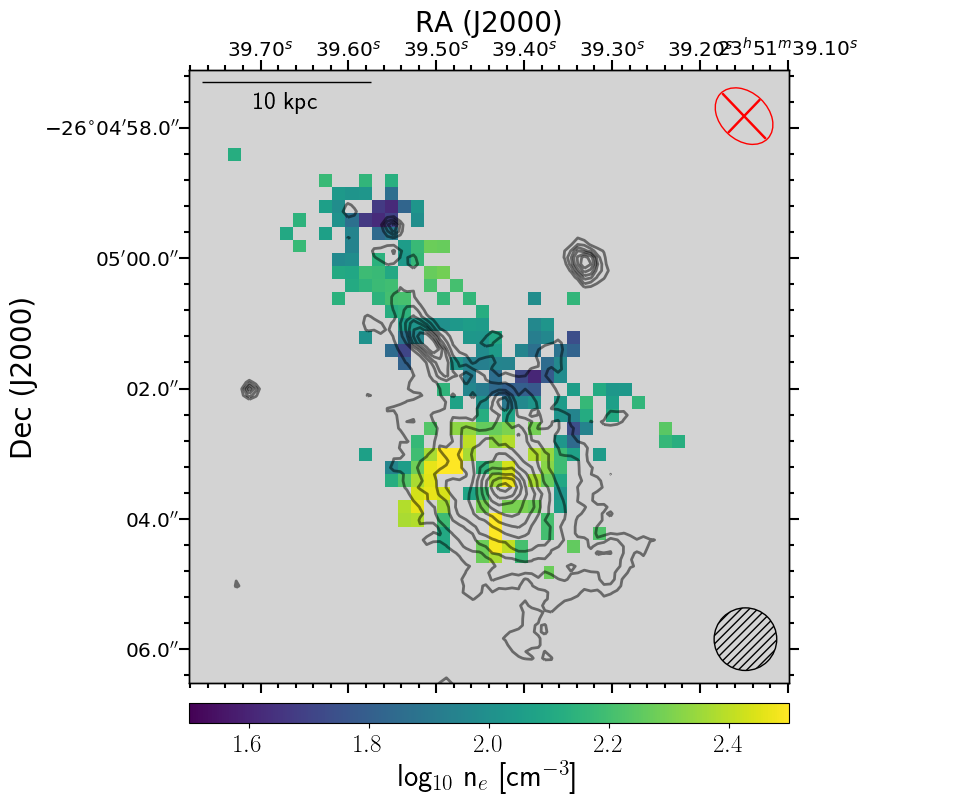}
    \includegraphics[width=\columnwidth]{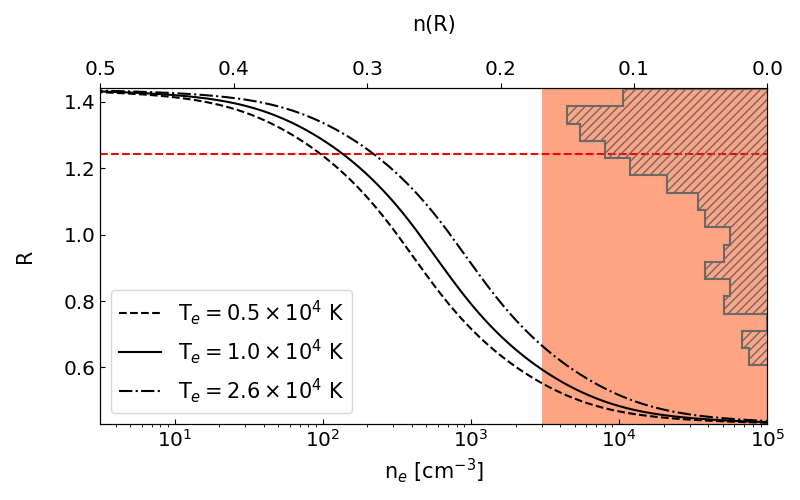}
    \caption{\textbf{Left panel}: Electron density (n$_e$) map for Abell 2667 BCG as retrieved from Eq.~\ref{eq:electron_density} for an electron temperature T$_e = 1.0\times 10^4$ K, with contours and symbols as defined in Fig.~\ref{fig:stars_vel_sigma}. \textbf{Right panel}: Electron density (n$_e$) curve \citep{2014A&A...561A..10P} as a function of the intensity ratio R = [SII]$\lambda6716/$[SII]$\lambda6731$ for T$_e = 0.5\times 10^4$ K (\textit{black dashed line}), T$_e = 1.0\times 10^4$ K (\textit{black solid line}) and T$_e = 2.6\times 10^4$ K (\textit{black dash-dotted line}). The red area of the diagram highlights the electron density values above the [SII] doublet critical density (i.e. n$_e \sim 3 \times 10^3$ \cmcub). We present the distribution of the retrieved R values as well as the median value of the distribution R$\sim 1.21$ (\textit{red dashed line}).}
    \label{fig:electron_density}
\end{figure*}

In Fig.~\ref{fig:electron_density} (left panel), the electron density (n$_e$) map of the gas is presented. 
The n$_e$ values were obtained from the intensity ratio R = [SII]$\lambda6716/$[SII]$\lambda6731$ according to the empirical equation by \cite{2014A&A...561A..10P}:
\begin{align}
\text{log}_{10}~ \text{n}_e [\text{cm}^{-3}] =&~0.0543~\text{tan}(-3.0553~\text{R}+2.8506) \nonumber\\
&+6.98 -10.6905~\text{R} +9.9186~\text{R}^2 \nonumber\\
&-3.5442~\text{R}^3 + 0.5~\text{log}_{10}\left[\frac{\text{T}_e [\text{K}]}{10^4}\right]
\label{eq:electron_density}
\end{align} 
and it was calculated supposing a spatially constant electron temperature (i.e. T$_e$) of $10^4$ K, a value widely adopted in the literature (e.g. \citealt{2015A&A...582A..63C}) if no direct electron temperature information can be recovered from the spectra (e.g. from the diagnostic diagrams based on [OIII]$\lambda\lambda(4958+5007)/\lambda 4363$ or [ArIII]$\lambda\lambda(7135+7751)/\lambda 5192$, see \citealt{2014A&A...561A..10P}).

A rather flat pattern for the electron density was retrieved, with a median value n$_e \sim 10^{2.2}$ \cmcub. 
Only the inner region of the BCG shows a higher value for n$_e$ although always lower than $3\times10^3$ \cmcub. 
These values are well below the critical density of the [SII] doublet, i.e. n$_e \sim 3\times 10^3$ \cmcub~(see \citealt{2006agna.book.....O}) above which the lines become collisionally de-excited. 
Therefore, we can consider reliable within the errors the electron density derived.
Nonetheless, to probe a possible dependence of the electron density from the electron temperature, we let vary T$_e$ within the range from $5000$ K to $26000$ K (i.e. the dashed and dash-dotted lines in Fig.~\ref{fig:electron_density} right panel, respectively). 
The median values for n$_e$ do not show substantial variations, going from $\sim 10^{2}$ \cmcub~(for T$_e = 5000$ K) to $\sim 10^{2.4}$ \cmcub~(for T$_e=26000$ K).

\section{Discussion and Conclusions}
\label{sec:conclusions}
In this paper, we have extensively analysed the properties of the BCG inhabiting the cool core cluster Abell 2667, based on both \textit{HST} imaging and MUSE data.

The bi-dimensional modelling of the galaxy surface brightness profile on the \textit{HST} images (F450W, F606W and F814W filters) with a S\'ersic law, has allowed us to probe the structural properties of the BCG.
The subtraction of the surface brightness model from the \textit{HST} observations has revealed a complex system of filamentary substructures all around the BCG and lying along the projected direction of the galaxy major-axis. 
These substructures appear to be constituted by clumps of different size and shape, plunged into a more diffuse component.
The \textit{HST} intense `blue' optical colour of these clumps is indicative of on-going star formation activity, as also corroborated by the spatially resolved current star formation rate we obtained from MUSE spectra.
Our BPT diagrams seem to further support this hypothesis since the emission arising from the clumps is `composite', thus requiring a contribution from both AGN and star formation.
The map of the galaxy current star formation rate, i.e. the SF of the last $0.02$~Gyr, shows a burst of activity with a spatially integrated rate of $\simeq 55~\text{M}_\odot~\text{yr}^{-1}$ extending along the same projected regions of the substructures. 
Our SFR is sensitively higher than the $8.7~\text{M}_\odot~\text{yr}^{-1}$ reported by \cite{2012ApJ...747...29R} and inferred via spectral energy distribution templates \citep{2009ApJ...692..556R} from the far infrared.
However, the difference between the two estimates is merely a consequence of the methodologies applied.
Both current SFRs can be considered as an upper and lower limit, respectively.
In fact, while the value inferred from the far infrared does not take into account the contribution from unobscured star formation, our estimate is contaminated by the AGN in the innermost regions of the galaxy.
To solve the problem, a new spectral energy distribution analysis of the galaxy is being carried on, taking into account a multi-wavelength set of data ranging from the ultraviolet to radio frequencies (Iani et al. \textit{in prep.}).

According to the galaxy S\'ersic index ($n = 4.64$) and its spatially resolved stellar kinematics, the BCG is a dispersion-supported spheroid with a circularised effective radius of $R_{e,circ}\simeq30$~kpc, a low and incoherent stellar line-of-sight velocity pattern ($\Delta$v$_\star \leq 150$~\kms) but high central velocity dispersion ($\sigma_0 \sim 216$~\kms).
Interestingly, the map of the stellar line-of-sight velocity reveals a redshifted pattern for all the spectra coinciding with the clumpy structures in the \textit{HST} filaments, thus suggesting these stars are gravitationally bounded together and, possibly, equally distant from the BCG.
The BCG radial profile of the stellar velocity dispersion suggests a positive gradient.
The presence of a positive gradient has been observed in a conspicuous number of central galaxies studied with IFU data (\citealt{2018MNRAS.477..335L}, \citealt{2018MNRAS.473.5446V}) even though its physical origin has not been understood yet.
Increasing velocity dispersion profiles are usually  interpreted as evidence that the diffuse stellar halo consists of accumulated debris of stars stripped from cluster members as a consequence of minor mergers ($\leq1:4$ mass ratios).
However, due to the low SNR of the external spaxels in the maps of Fig.~\ref{fig:stars_vel_sigma} and the presence of possible artefacts, our considerations on the velocity dispersion profile are limited to the most inner regions of the galaxy ($r \leq 10$~kpc).

Our estimate of the BCG stellar mass is $\sim 1.38\times 10^{11}~\text{M}_\odot$.
This value takes into account both stars in the nuclear-burning-phase and remnants.
In line with our findings on the galaxy mass assembly history, the $57.2$\% of the galaxy stellar mass was assembled more than $5.6$~Gyr ago  ($z>0.7$) while in later times episodes of star formation seem to have formed the remaining $39.1$\% ($0.6<t_{lb}<5.6$~Gyr), $3.1$\% ($0.02<t_{lb}<0.6$~Gyr), $0.6$\% ($t_{lb}<0.02$~Gyr). 
As shown in previous works (\citealt{2004MNRAS.347..740K}, \citealt{2015MNRAS.449.3347O}), this suggests that the BCG mass growth at high redshift ($z\geq 0.8$) has been principally driven by monolithic collapse and major mergers ($\geq1:3$ mass ratios) while at lower redshift minor mergers and cold accretion could have contributed the most to the stellar mass growth.

Our target galaxy is known to host a radio-loud Type 2 AGN (\citealt{2013MNRAS.432..530R}, \citealt{2015A&A...581A..23K}, \citealt{2018ApJ...859...65Y}).
Thus, applying the $M_{SMBH} - \sigma_0$ relation by \cite{2012MNRAS.426.2751Z}, we estimate the mass of the SMBH hosted by the BCG equal to $3.8^{+1.7}_{-1.3}\times 10^9~\text{M}_\odot$.
In agreement with the results obtained by \cite{2009ApJ...699L..43S} on a large survey of AGNs, the SMBH mass obtained is typical of low excitation galaxies. 
This is also confirmed by our BPT diagnostic diagrams, which classify the emission coming from the BCG core as LINER.
In the BCG's outskirts, the line ratios classify the observed emission as `composite', suggesting the possible presence of both star formation and an attenuated UV radiation field from the AGN.
Similar results have been observed by \cite{2007MNRAS.380...33H} for a sample of X-ray bright BCGs residing in cool core clusters at lower redshifts ($z<0.2$), thus showing the complexity of the phenomena arising line-emitting nebul\ae~around BCGs.

Ultimately, from the analysis of the integrated flux of the \hb~and \oiii~emission lines within different velocity channels, we detect the presence of two decoupled gas components with velocities in between $\Delta$v$_{gas} = [+100;+500]$~\kms~and $\Delta$v$_{gas} = [-500;-100]$~\kms.
The two gas components appear to cover different spatial extents, with the blueshifted stream more compact and centred around the galaxy core while the redshifted flow spreads along the regions hosting the \textit{HST} substructures.
This result shows that the gas and stars in the filaments are comoving and redshifted with respect to the observer. 

To account for both the filamentary structures and the two gas components observed, two different scenarios can be invoked: a model of ICM accretion with AGN feedback and a merger.
The first scenario is supported by the fact that Abell 2667 is a cool core cluster ($t_{cool} = 0.5$~Gyr, \citealt{2009ApJS..182...12C}) and therefore, we expect the presence of an inflow of cold ICM towards the BCG.
In this case, according to the \textit{chaotic cold accretion model} (e.g. \citealt{2017ApJ...837..149G}; \citealt{2016Natur.534..218T}; \citealt{2018ApJ...865...13T}) and as predicted by recent theory (e.g. \citealt{2005ApJ...632..821P}; \citealt{2015Natur.519..203V}; \citealt{2015ApJ...808L..30V}) and simulations (e.g. \citealt{2012MNRAS.420.3174S}, \citealt{2013MNRAS.432.3401G}, \citealt{2014ApJ...789..153L}), the inflow of cold ICM would be a stochastic and clumpy event giving rise to cold giant clouds.
Within these clouds, the physical conditions of the gas could satisfy the criteria for the ignition of star formation, thus explaining the observed diffuse and clumpy star-forming filaments revealed by \textit{HST}.
Since the filamentary structures extend down to the very centre of the BCG, this inflow of material could be at the origin of the galaxy AGN activity.
The AGN ignition by the inflow could also be responsible for feedback through outflows that could in turn regulate the SMBH accretion in agreement with the theoretical models introduced to overcome the \textit{cooling flow problem} (e.g. \citealt{2007ARA&A..45..117M}; \citealt{2012ARA&A..50..455F}; \citealt{2013MNRAS.432.3401G}). 
In this scenario, the redshifted stream of gas we detect would relate to the infalling material and clouds from the ICM while the blueshifted flow would be compatible with a massive ionised gas outflow originating from the SMBH.
Assume a bi-conical emission, the redshifted component of the nuclear outflow could be not revealed due to absorption from the galactic medium.
An alternative flavour for this scenario, resembling to what found in recent studies (e.g. \citealt{2017Natur.544..202M}; \citealt{2018ApJ...865...13T}), could be taken into account too. 
In this case, the AGN feedback would be the mechanism triggering, influencing and regulating the star formation within the filamentary substructures detected with \textit{HST}.
This scenario would happen if the observed substructures were made of compressed gas along possible AGN jets.
Even though the physical scale at which we observe the clumps ($\sim 10-20$~kpc from the galaxy nucleus) is comparable with the above-mentioned works, we tend to exclude this hypothesis since our preliminary analysis of the public radio data-set (JVLA, ALMA), as well as the X-rays (\textit{Chandra}), of the galaxy seem to rule out the presence of jets or inflated bubbles on such a large scale.

The second scenario envisages the Abell 2667 BCG merging with a small disc-like galaxy.
In this case, an infalling distorted disc of small size, coinciding with the biggest clump in the filaments, is tidally disrupted and is leaving behind streams of gas that are triggering star formation (i.e. the smaller knots) possibly because of the interaction with the BCG circumgalactic medium.
Supporting this scenario, the clumps appear to lie on the BCG galactic plane, as it would happen with an accreting satellite galaxy that loses its angular momentum and starts inspiraling towards the BCG centre. 
At the opposite extremity of the galaxy with respect to the clumps in the \textit{HST} images, diffuse blue emission is present likely due to the gas left behind the satellite galaxy from a previous orbit.

We are planning a follow up of this source in both X-rays and the radio domain since deeper observations in both electromagnetic bands will be crucial to obtain detailed information about the X-ray emission and the molecular cold gas component of this source, allowing us to disentangle the imprints of the evolutionary scenarios proposed.

\section*{Acknowledgements}
We thank the anonymous referee for useful comments and a careful reading of the manuscript.
This work is based on observations collected at the European Southern Observatory under ESO programmes 094.A-0115(A) and on observations made with the NASA/ESA Hubble Space Telescope, and obtained from the Hubble Legacy Archive, which is a collaboration between the Space Telescope Science Institute (STScI/NASA), the Space Telescope European Coordinating Facility (ST-ECF/ESA) and the Canadian Astronomy Data Centre (CADC/NRC/CSA). 
GR and CM acknowledge support from an INAF PRIN-SKA 2017 grant.
AM acknowledges funding from the INAF PRIN-SKA 2017 program 1.05.01.88.04.
PGP-PG acknowledges support from the Spanish Government grant AYA2015-63650-P.
We warmly thank Stefano Ciroi and Bianca Maria Poggianti for helpful discussion.




\bibliographystyle{mnras}
\bibliography{biblio.bib} 



\appendix
\section{Abell 2667 BCG surface brightness radial profile}
\begin{figure*}
    \centering
    \includegraphics[width=0.7\textwidth]{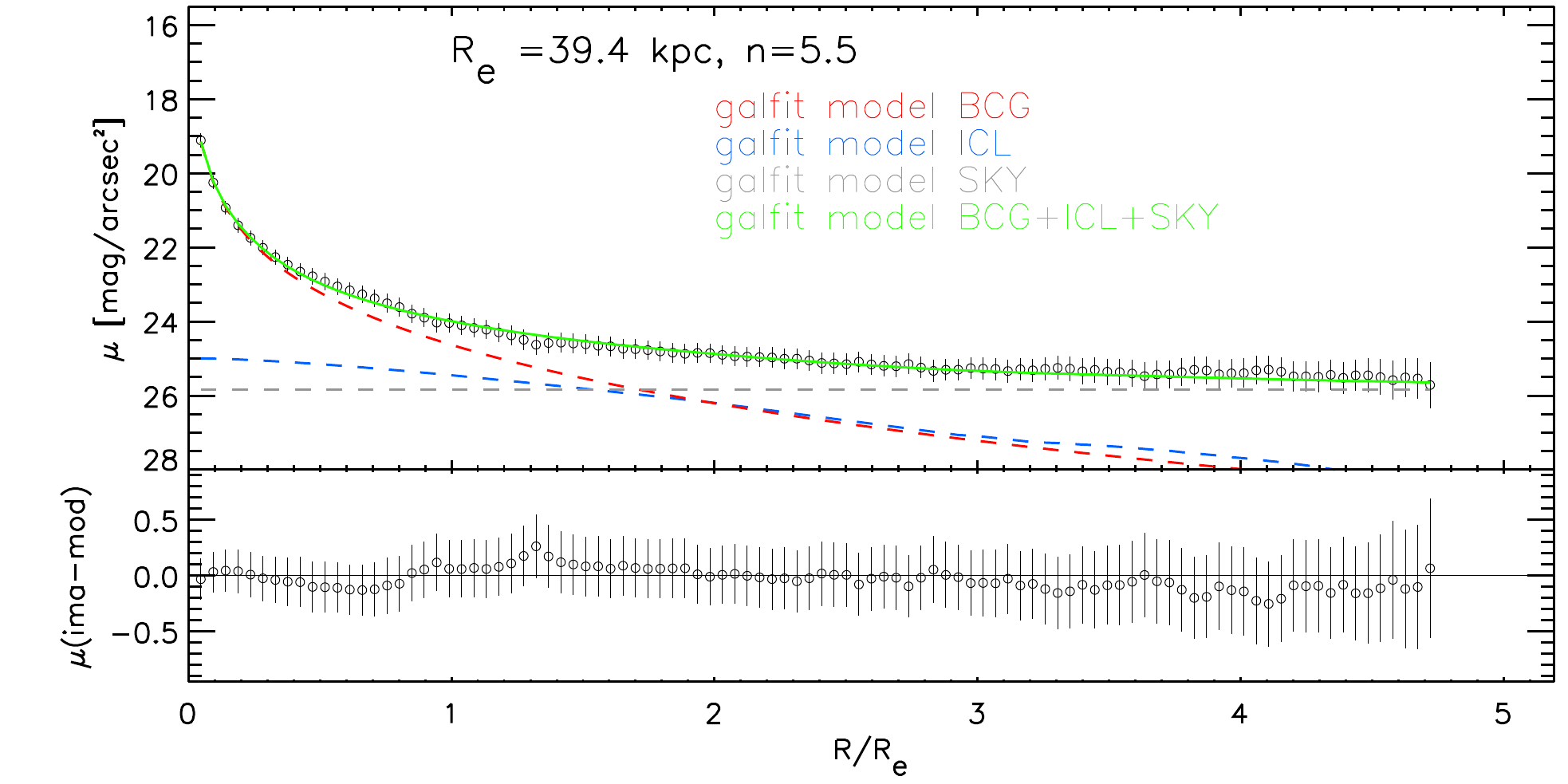}
    \caption{\textit{galfitm} fit (\textit{green solid line}) of the BCG surface brightness profile in the F814W filter. The dashed lines highlight the best matching radial profiles for the sky background (\textit{grey line}), the ICL (\textit{blue line}) and the central galaxy (\textit{red line}).}
    \label{fig:1d_surf_b_prof}
\end{figure*}

In Fig.~\ref{fig:1d_surf_b_prof} we present the 1D radial profile of the observed Abell 2667 BCG surface brightness as extracted from the HST F814W band image.
We also show the best matching \textit{galfitm} components for the sky background, intracluster light and galaxy emission (i.e. S\'ersic profile).
Details on the fitting procedure are presented in Sec.~\ref{sec:galfit}.

\section{MUSE integrated spectrum of Abell 2667 BCG}
\begin{figure*}
    \centering
    \includegraphics[width=\textwidth]{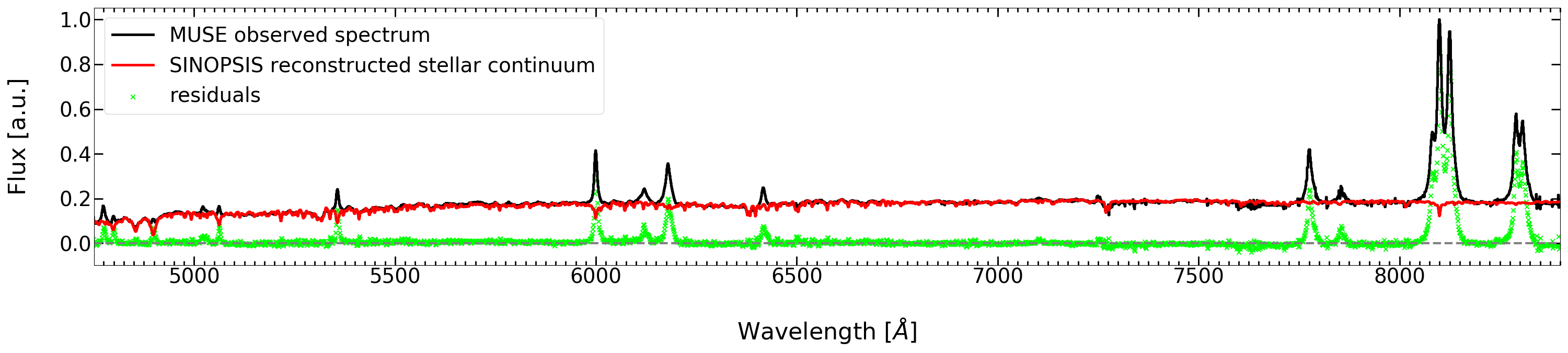}
    \caption{Example of a MUSE observed spectrum (\textit{black solid line}) and its SINOPSIS reconstructed stellar continuum (\textit{red solid line}) of the Abell 2667 BCG. The green crosses show the residuals between the observed and reconstructed model, i.e. the spectrum of only emission lines that we use for the analysis of the gas component in Sec.~\ref{subsec:gas_component}.}
    \label{fig:gal_spec}
\end{figure*}

In Fig.~\ref{fig:gal_spec} we show the MUSE integrated spectrum of the Abell 2667 BCG. 
We also present the stellar continuum reconstructed by SINOPSIS and the residuals between the observed spectrum and the model.

\section{detection of a Ly$\alpha$-emitter candidate}
\begin{figure*}
    \centering
    \includegraphics[width=\textwidth]{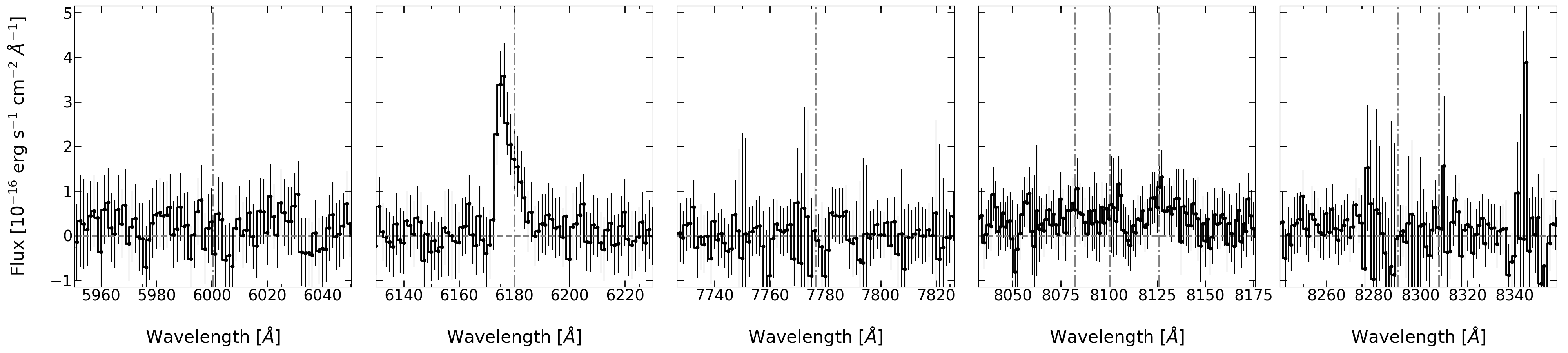}
    \caption{Integrated spectrum of the Ly$\alpha$-emitter candidate at $z \sim 4.08$ (F region, see Fig.~\ref{fig:hb_oiii_fluxes}). We present wavelength cutouts of the MUSE observed spectrum (\textit{grey solid line}) and its 3-$\sigma$ error, integrated over a $1.2'' \times 1.2''$ region centred at RA(J2000.0)$=23^h51^m39.507^s$, Dec(J2000.0)$=-26^\circ04'56.81''$. In the cutouts, the vertical dash-dotted lines indicate the expected wavelength of \hb, [OIII]$\lambda5007$, [OI]$\lambda6300$, \halpha+[NII] and the [SII] doublet at the BCG redshift. The only line we detect falls in the [OIII]$\lambda5007$ cutout. However, the absence of the weaker component of the [OIII] doublet and of the other lines, together with the highly asymmetric shape of the emission, suggests that the detection corresponds to a different line, possibly a Ly$\alpha$.}
    \label{fig:lya_spec}
\end{figure*}

In Fig.~\ref{fig:lya_spec} we present wavelength cutouts of the MUSE integrated spectrum for the Ly$\alpha$-emitter candidate at $z \sim 4.08$ (see Sec.~\ref{subsec:velocity_channels}).

\section{Affiliations}
\textit{
$^1$Dipartimento di Fisica ed Astronomia, Universit\`a degli Studi di Padova, Vicolo dell'Osservatorio 3, I-35122 Padova, Italy\\
$^{2}$European Southern Observatory, Karl Schwarzschild Stra\ss e 2, D-85748 Garching, Germany\\
$^3$INAF - Osservatorio Astronomico di Padova, Vicolo dell'Osservatorio 5, I-35122 Padova, Italy\\
$^4$Instituto de Radioastronom\'ia y Astrof\'isica, UNAM, Campus Morelia, A.P. 3-72, C.P. 58089, Mexico\\
$^5$INAF - Osservatorio Astrofisico di Arcetri, Largo Enrico Fermi 5, I-50125 Firenze, Italy\\
$^6$Dipartimento di Fisica e Scienze della Terra, Universit\`a degli Studi di Ferrara, Via Saragat 1, I-44122 Ferrara, Italy\\
$^7$Kapteyn Astronomical Institute, University of Groningen, Postbus 800, 9700 AV Groningen, The Netherland\\
$^8$Institut de Radioastronomie Millim\'etrique 300 rue de la Piscine, Domaine Universitaire 38406 Saint Martin d'H\`eres, France\\
$^{9}$Excellence Cluster Universe, Boltzmannstra\ss e 2, D-85748 Garching, Germany\\
$^{10}$SOFIA Science Center, USRA, NASA Ames Research Center, M.S. N232-12 Moffett Field, CA 9403\\
$^{11}$INAF - Osservatorio Astronomico di Capodimonte, via Moiariello 16, 80131 Napoli, Italy\\
$^{12}$Departamento de Astronom\'ia y Astrof\'isica, Universidad Complutense de Madrid, Av. Complutense s/n, C.P. 28040, Madrid, Spain\\
$^{13}$Centro de Astrobiolog\'{\i}a (CAB, INTA-CSIC), Carretera de Ajalvir km 4, E-28850 Torrej\'on de Ardoz, Madrid, Spain\\
$^{14}$Dipartimento di Fisica e Astronomia, Universit\`a degli Studi di Bologna, via Gobetti 93/2, I-40129 Bologna, Italy\\
$^{15}$INAF - Istituto di Radioastronomia - Italian node of the ALMA Regional Centre (ARC), via Gobetti 101, I-40129 Bologna, Italy\\
$^{16}$Departamento de Astronom\'ia, Universidad de Concepci\'on, Barrio Universitario, Concepci\'on, Chile\\
$^{17}$Leiden Observatory, Leiden University, PO Box 9513, 2300 RA Leiden, The Netherlands\\
}

\bsp	
\label{lastpage}
\end{document}